\documentclass{aa}

\usepackage{graphicx}
\usepackage{amssymb,amsmath}
\usepackage{xspace}
\usepackage[english]{babel}
\usepackage{hyperref}
\usepackage{amsmath}
\usepackage{longtable}
\usepackage{comment}
\usepackage{booktabs}
\addto\extrasenglish{%
}
\usepackage{txfonts}
\usepackage{dblfloatfix}

\newcommand{\eg}{\emph{e.g.}}
\newcommand{\ie}{\emph{i.e.}}

\begin{document}

   \title{Validation of the \citet{2010ApJ...716....1B} SDSS-derived kinematic models for the Milky Way's disk and halo stars with \textit{Gaia} Data Release 3 proper motion and radial velocity data}

   % \subtitle{I. Overviewing the $\kappa$-mechanism}

   \author{Bruno Dom\'{i}nguez\inst{1}\fnmsep\thanks{The three authors contributed equally}
          \and
          Siddharth Chaini\inst{2}\fnmsep$^\star$
          \and
          Karlo Mrakov\v{c}i\'{c}\inst{3}\fnmsep$^\star$
          \and
          Brandon Sallee\inst{4}
          \and
          \v{Z}eljko  Ivezi\'{c}\inst{5}\fnmsep\thanks{Corresponding author -- \email{ivezic@uw.edu}}
          }

   \institute{Departamento de Astronom\'{i}a, Facultad de Ciencias,                Universidad de la Rep\'{u}blica, Igu\'{a} 4225, 11400,               Montevideo, Uruguay\
                % \email{bdominguez@fcien.edu.uy}
        \and
             Department of Physics and Astronomy, University of Delaware, Newark, DE 19716, USA\
        \and
            Faculty of Physics, University of Rijeka, Radmile Matej\v{c}i\'{c} 2, 51000 Rijeka, Croatia\
        \and
            Department of Astronomy, University of Washington, Box 351580, Seattle, WA 98195, USA\
        \and
            Department of Astronomy and the DiRAC Institute,
            University of Washington, 3910 15th Avenue NE, Seattle, WA 98195, USA
             }

   \date{Received \today; accepted XXX}

  \abstract
   {We validate the \citet{2010ApJ...716....1B}
kinematic models for the Milky Way's disk and halo stars with \textit{Gaia} Data Release 3 data.
Bond et al. constructed models for stellar velocity distributions using stellar radial velocities measured by the Sloan Digital Sky Survey (SDSS) and stellar proper motions derived from SDSS and the Palomar Observatory Sky Survey astrometric measurements. These models describe velocity distributions as functions of position in the Galaxy, with separate models for
disk and halo stars that were labeled using SDSS photometric and spectroscopic metallicity measurements.
We find that the Bond et al. model predictions are in good agreement with recent measurements of stellar radial velocities
and proper motions by the \textit{Gaia} survey. In particular, the model accurately predicts the skewed non-Gaussian distribution of rotational velocity for disk
stars and its vertical gradient, as well as the dispersions for all three velocity components.
Additionally, the spatial invariance of velocity ellipsoid for halo stars when expressed in spherical coordinates is also confirmed by \textit{Gaia}
data at galacto-centric radial distances of up to 15 kpc. }


    \titlerunning{Validation of the \citet{2010ApJ...716....1B} models with \textit{Gaia} DR3 data}
    \authorrunning{B. Dom\'{i}nguez et al.}
    
   \maketitle
%
%-------------------------------------------------------------------

\section{Introduction\label{sec:intro}}

The Milky Way is a complex and dynamic structure that is constantly being shaped by the infall of matter from the Local
Group and mergers with neighboring galaxies. The advent of modern sky surveys has resulted in a wealth of data for a
large number of individual stars, enabling their distribution to be described and modeled in a nine-dimensional space encompassing three spatial coordinates, three velocity components, and three principal stellar parameters: luminosity, effective temperature, and metallicity \citep[for a detailed discussion and references see, \eg,][]{2012ARA&A..50..251I}.
The modeling of observed multi-dimensional distributions can provide in-situ constraints on the formation and evolutionary
processes of a spiral galaxy, such as the Milky Way. For example, stellar number density and kinematic measurements, aided by
models, can be used to constrain the Milky Way's dark matter distribution \citep[\eg,][]{2012ApJ...758L..23L, 2014ApJ...794..151L}.

Stellar distribution models also play a crucial role in predicting stellar content and its expected behavior in upcoming surveys, such as Rubin's Legacy Survey of Space and Time \citep[LSST;][]{2019ApJ...873..111I}. 
For example, \cite{2022ApJS..262...22D} have recently produced a massive simulated stellar catalog\footnote{This catalog
is publicly available from NOIRLab's Astro Data Lab portal.} corresponding to the anticipated sky coverage and depth
of LSST ($r\sim27$). In addition to this catalog's current use in quantitative forecasting of LSST science outcome, such catalogs will also become
increasingly important as inputs for tools that quantify various LSST selection functions (\eg, photometric selection of quasars, where
stars are rejected using color and proper motion cuts) and for providing Bayesian priors for science analysis (\eg, for stellar
photometric distance estimation). The primary objective of this paper is to validate the kinematic models that can be used when generating such catalogs. 

\cite{2010ApJ...716....1B} studied Milky Way kinematics using a sample of about 19 million main-sequence stars with Sloan Digital
Sky Survey (SDSS) photometry and proper-motion measurements derived from both the SDSS and the Palomar Observatory Sky
Survey (POSS) astrometry, including a subset of $\sim$170,000 stars with radial-velocity measurements from the SDSS spectroscopic survey.
They developed a simple descriptive model for the overall kinematic behavior that captures these features over most of the probed
volume (summarized in Sect. \ref{sec:BondModels}), with distances in the range from 100 pc to 10 kpc and over a quarter of the sky
at high Galactic latitudes ($|b| > 20^\circ$).

\textit{Gaia's} recently published photometric, radial velocity and proper motion measurements \citep{2021A&A...649A...1G} enable us to critically assess the accuracy of \cite{2010ApJ...716....1B} kinematic models because of their superior uncertainties
(for a comparison of SDSS and SDSS-POSS measurement uncertainties with \textit{Gaia's}, see Figure 21 in \citealt{2012ARA&A..50..251I}). 
In addition, with \textit{Gaia's} data, we can extend the model validity to distances below 100 pc (due to SDSS image saturation at $r\sim14$),
and to a four times larger full-sky area. We describe our datasets and analysis methodology in Sect. \ref{sec:analysis}, and we present our analysis results in
Sect. \ref{sec:results}. Our main results are summarized and discussed in Sect. \ref{sec:disc}. All our results from this paper can be reproduced from our accompanying GitHub repository.\footnote{\url{https://github.com/sidchaini/MWKinematicsFGKM}}

%--------------------------------------------------------------------
\section{Data description and analysis methodology\label{sec:analysis}}

\subsection{Stellar sample selection}

\begin{figure}[ht]
%\hskip 0.01in
\centering
\includegraphics[width=0.999\linewidth,angle=0]{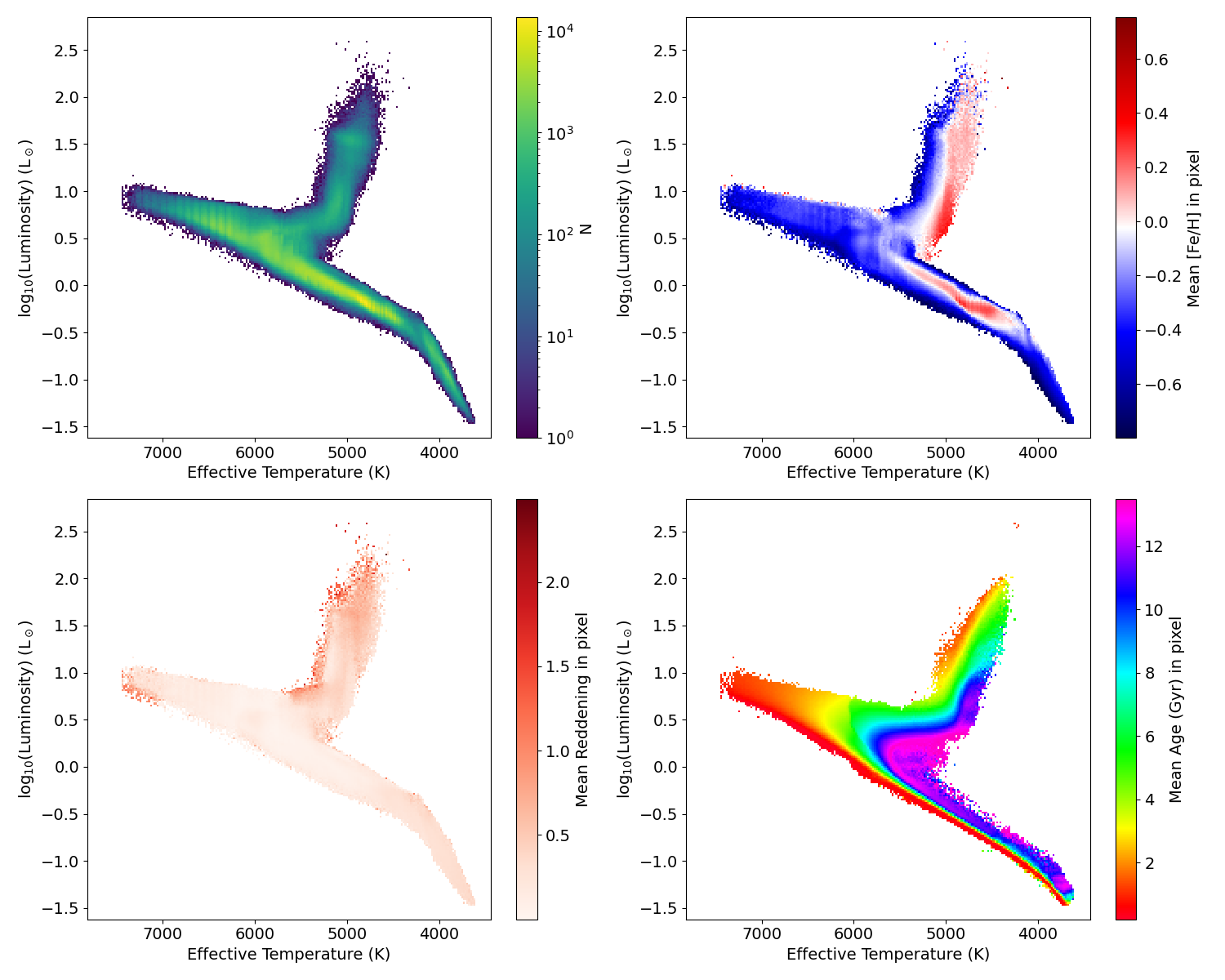}
\caption{Distribution of the 3.3 million FGKM stars discussed here in the Hertzsprung-Russell diagram.
The top left panel shows stellar counts on a log scale and the remaining panels show mean per-pixel values of metallicity, interstellar dust
reddening and age. Note that the hottest stars in the sample (effective temperature above 6,500 K) are also the youngest (ages below 2 Gyr).} 
\label{fig:HRdiagram} 
\end{figure}

Blue main sequence stars close to the turn-off point are ideal for studying bulk kinematic properties in the Milky
Way because they are numerous, have comparatively large limiting distances due to high luminosities, and their metallicities can be
sufficiently accurately (0.1 -- 0.2 dex) estimated from broad-band photometric data \citep{2008ApJ...684..287I}. 
These stars were principal stellar populations used for developing a stellar number density model in \cite{2008ApJ...673..864J} and stellar
kinematics models in \cite{2010ApJ...716....1B}. While \textit{Gaia} Data Release 3 data access tools could be used to construct an adequate
sample of stars for comparison to SDSS results, an essentially perfect catalog for this purpose was recently published by the \textit{Gaia team}.

Creevey, Sarro, Lobel et al. \citep{2023A&A...674A..39G} constructed a carefully cleaned science-ready sample (``a golden sample of
astrophysical parameters'') of F, G, K, and M stars listed in the \textit{Gaia} Data Release 3 catalog. After quality cuts based on astrometric,
photometric, and astrophysical parameters, along with other Gaia-based criteria, they provided a catalog of 3.3 million stars with
measured and/or estimated astrophysical parameters using \textit{Gaia} and other datasets (including effective temperature, metallicity, mass,
age and spectral type). Of particular relevance to the analysis presented here, their FGKM sample included the following selection criteria:
an effective signal-to-noise ratio cut applied through a limit on the $B_P-R_P$ color error $\sigma < 0.06$ mag, effective temperature
$T_{eff} < 7500$ K, absolute magnitude in \textit{Gaia's} G band $M_G < 12$, and metallicity [Fe/H]$ >-0.8$. Consequently, the sample is
dominated by high-metallicity disk stars at distances of up to about 1 kpc (the median distance is 620 pc).
In subsequent analysis, we use stellar distances computed by \cite{2021AJ....161..147B}. 

\begin{figure}[ht]
\centering
%\hskip 0.01in
\includegraphics[width=0.999\linewidth,angle=0]{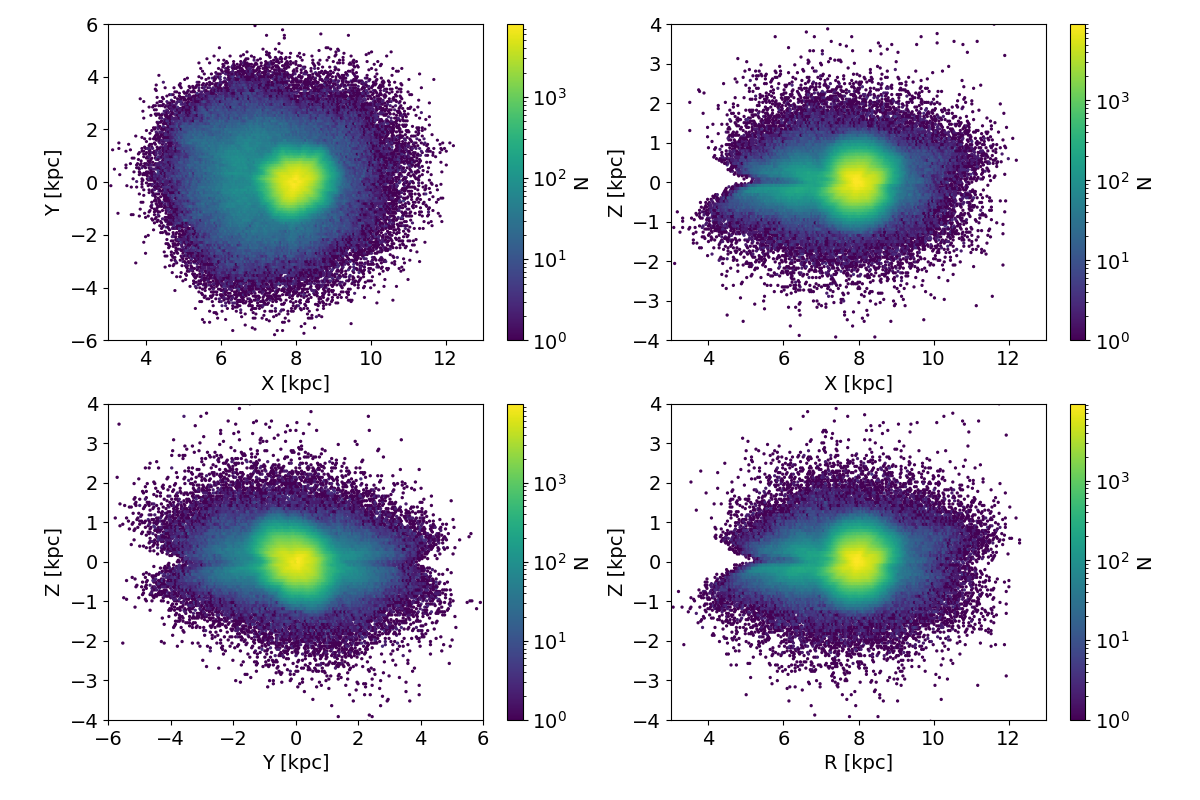}
\caption{Distribution of the stars shown in Figure~\ref{fig:HRdiagram} in the right-handed galactocentric Cartesian coordinates ($R_\sun=8.0$
kpc). Note that the counts are shown on a log scale -- the majority of stars are at distances below 1 kpc (the median distance is 620 pc).}
\label{fig:XYZfgkm}
\end{figure}

\begin{figure*} [h]
\centering
%\hskip 0.01in
\includegraphics[width=0.999\linewidth,angle=0]{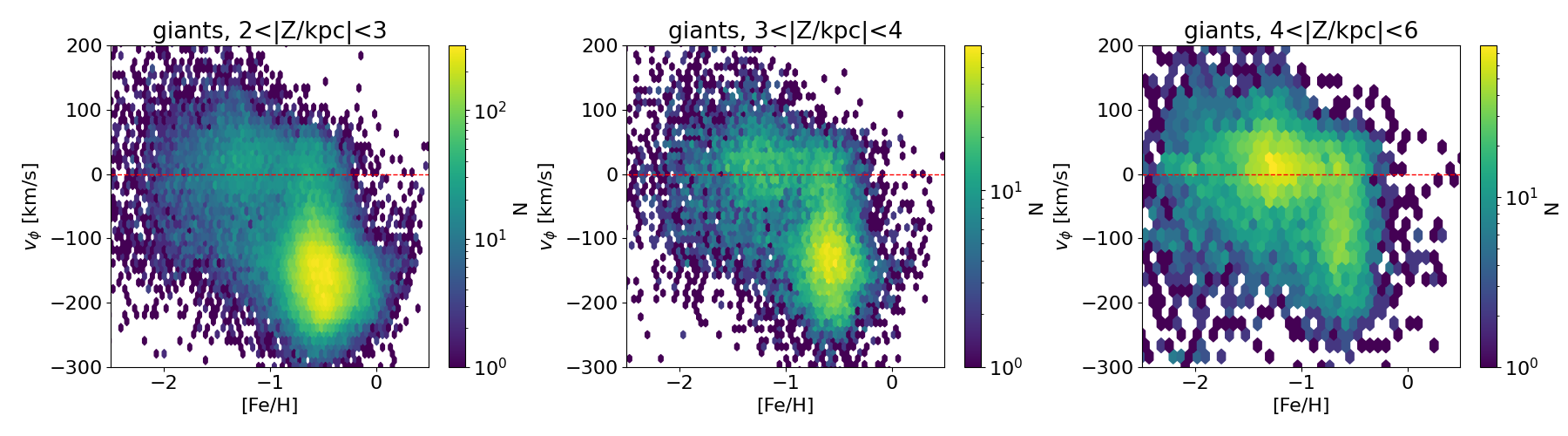}
\caption{Rotational velocity vs. metallicity distribution of the red giant stars from the \cite{2023ApJS..267....8A}
catalog that have a cylindrical galacto-centric radius between 7 kpc and 10 kpc. Each panel are different bins of distance
from the Galactic plane (from left to right: 2 -- 3 kpc, 3 -- 4 kpc, 4 -- 6 kpc). The mean rotational velocity for halo stars is
consistent with 0, with metallicity $[Fe/H] < -1$, while for disk stars $[Fe/H] > -1$ and rotation slows down with
distance from the plane. }  
\label{fig:FeHZrg}
\end{figure*}

\autoref{fig:HRdiagram} shows the distribution of this sample (hereafter the FGKM sample) in the Hertzsprung-Russell diagram, and also
provides information about metallicity, interstellar dust reddening, and age distributions. Given distances, we computed
galactocentric coordinates for all stars using eqs.~15--16 (with $R_\sun=8.0$
kpc) from \cite{2008ApJ...673..864J}. The spatial distribution of this sample is
shown in \autoref{fig:XYZfgkm}. For 1.73 million stars in this sample, radial velocities are also available. We computed their
velocities in the cylindrical coordinate system using eqs.~4--8 from \cite{2010ApJ...716....1B}, using the same values of solar peculiar motion taken from \cite{1998MNRAS.298..387D}. These velocities, as well as proper
motions for all the stars, are used for model validation in the next section. 

\begin{figure}[h]
\centering
%\hskip 0.01in
\includegraphics[width=\linewidth,angle=0]{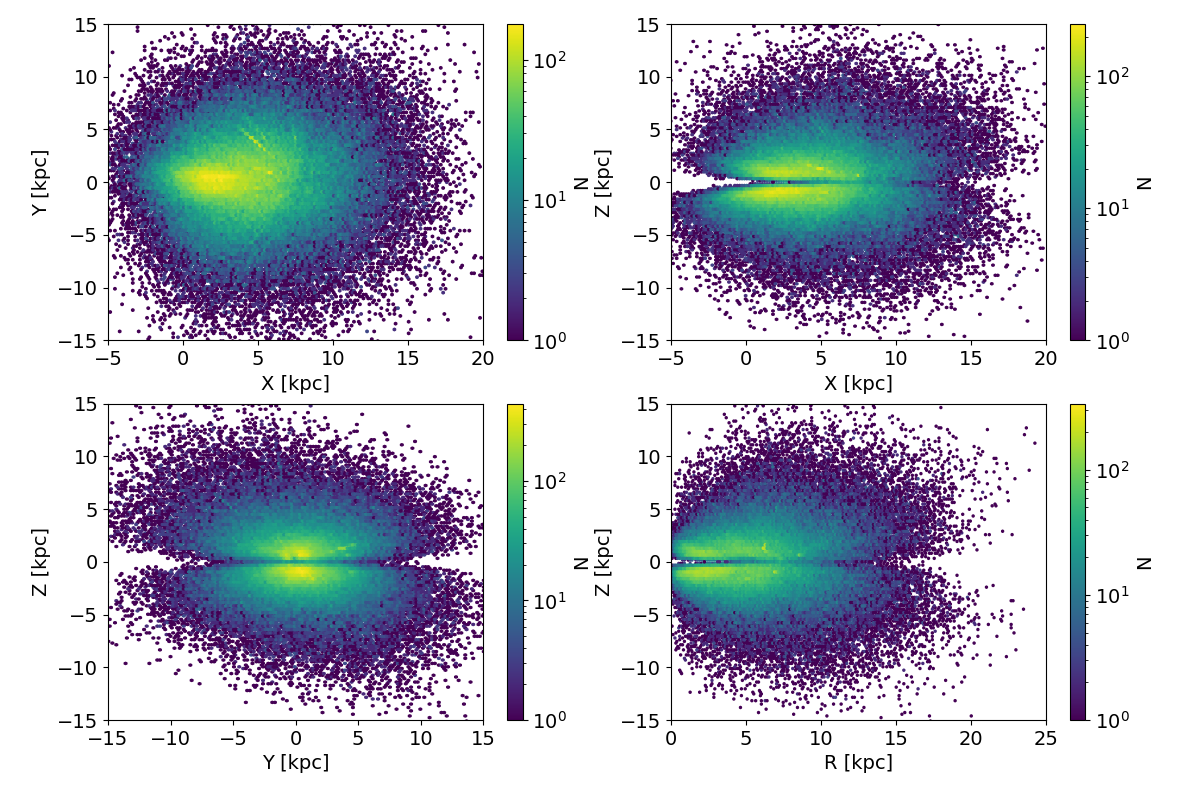}
\caption{Distribution of the 188,807 candidate halo red giant stars with [Fe/H]$<-1.2$. Their median
  distance is 6.0 kpc. Note the significantly expanded axis limits compared to Figure~\ref{fig:XYZfgkm}. }
\label{fig:XYZhalo}
\end{figure}

In addition to the FGKM sample with a median distance of 620 pc, we utilize another sample to analyze more distant disk stars and low-metallicity halo stars. \cite{2023ApJS..267....8A}
published a catalog of over 17 million bright red giants ($G < 16$), including metallicity estimates from \textit{Gaia's} XP spectra.
Figure~\ref{fig:FeHZrg} shows the rotational velocity vs. metallicity distribution for 2.5 million stars from the solar cylinder
(\ie, having a galacto-centric cylindrical radius between 7 kpc and 10 kpc), as a function of distance from the Galactic plane. As evident, 
metallicity can be used to separate non-rotating low-metallicity halo stars from rotating high-metallicity disk stars. 
Of the 283,616 red giants with [Fe/H] $<-1.2$ in the full sample, 188,807 have valid radial velocity measurements  
(that are not ``NaN''). Hereafter, we refer to the latter sample as the ``halo sample''; its spatial distribution
in the Galaxy is shown in \autoref{fig:XYZhalo}.

\subsection{Kinematic models derived from SDSS data\label{sec:BondModels}}

\cite{2010ApJ...716....1B} studied Milky Way kinematics using a sample of 18.8 million main-sequence stars with SDSS magnitude
$14<r<20$ and proper-motion measurements derived from both SDSS and POSS astrometry over a quarter of the sky at high Galactic latitudes
($|b| > 20^\circ$). Their sample also included $\sim$170,000 stars with radial-velocity measurements from the SDSS spectroscopic
survey. Distances to stars, in the range from 100 pc to 10 kpc, were determined using a color-based and metallicity-dependent
photometric parallax relation \citep{2008ApJ...673..864J, 2008ApJ...684..287I}.

% {\bf 
Once kinematic models, described in the following sections, are formulated in terms of physical quantities, 
velocities and distances, the systematic and random uncertainties for proper motion, radial velocity and distance estimates from \cite{2010ApJ...716....1B} would only matter in case of significant discrepancies between these models and \textit{Gaia} data. While we show below this not to be the case, we note that these uncertainties are quite well understood and discussed in detail in \cite{2010ApJ...716....1B}. 
Furthermore, an additional comparison of Gaia-based distances from  \cite{2021AJ....161..147B}
and SDSS-based photometric distances computed as in \cite{2010ApJ...716....1B} shows that the
two distance scales for high-metallicity disks stars agree at the level of a few percent, with a random scatter per star of about 15\% (L.~Palaversa, priv. comm.). We note that the analysis of kinematics by \cite{2010ApJ...716....1B} in their Section 5.4 limits distance-scale error for more distant and fainter low-metallicity halo stars to 5\%.
% }
% LP: agreement photom Mr with MrPi0: ['-0.013', '0.29', 17733]

\subsubsection{Kinematic model for disk stars\label{sec:BondModelsDisk}}

For disk stars, they found that in the region defined by 1 kpc $< |Z| < 5$ kpc, where $|Z|$ is the distance from the Galactic
plane, the rotational velocity for disk stars smoothly decreases, and all three components of the velocity dispersion increase, with $|Z|$.
They developed a simple descriptive model that effectively captures the overall kinematic behavior over most of the probed volume, as follows.

The decrease of rotational velocity with $|Z|$ for disk stars (often referred to as asymmetric drift, velocity lag, or velocity shear; for
more details and references to related work, see Sect.  3.4 in \citealt{2008ApJ...684..287I}) in the $|Z|=1-4$~kpc range can be described by
\begin{equation} 
\langle v_\phi \rangle = -205 + 19.2 \,\left|\frac{Z}{\rm kpc}\right|^{1.25} \,\,\, {\rm km~s^{-1}}.
\label{eq:Dlag}
\end{equation} 
The measured rotational velocity dispersion of disk stars increases with $Z$ faster than can be attributed to measurement errors;
their {\it intrinsic} velocity dispersion can be fit as
\begin{equation}
    \sigma_{\phi} = 30 + 3.0 \,\left|\frac{Z}{\rm kpc}\right|^{2.0} \,\,\, {\rm km~s^{-1}}.
\label{eq:Ddisp}
\end{equation} 
The errors on the power-law exponents of eqs.~\ref{eq:Dlag} and \ref{eq:Ddisp} are $\sim$0.1 and $\sim$0.2, respectively.

However, it was also found that a description of the full velocity distribution based solely on the first and second moments
(that is, by eqs.~\ref{eq:Dlag} and \ref{eq:Ddisp}) does not fully capture the detailed data behavior.  The observed non-Gaussian
shape for the rotational ($v_\phi$) component can be modeled by a sum of two Gaussians, with a fixed normalization ratio and a fixed
offset of their mean values for $\left|Z\right|<5$~kpc,
\begin{eqnarray} 
\label{eq:pDvPhi}
           p(x=v_\phi|Z)=f_1\,G[x|v_n(Z), \sigma_1] + f_2\,G[x|v_n(Z)  - \Delta v_n, \sigma_2], 
\end{eqnarray} 
where $G(x|\mu, \sigma)$ is Gaussian (normal) distribution, $f_1=0.75$, $f_2=0.25$, $\Delta v_n = 34$ ${\rm km~s^{-1}}$, and
velocity variation with $|Z|$ is captured by
\begin{equation} 
      v_n(Z) = v_0 + 19.2\,\left|\frac{Z}{\rm kpc}\right|^{1.25} \,\,\, {\rm km~s^{-1},} 
\label{eq:vnZ}
\end{equation}
with $v_0$ = $-194$ ${\rm km~s^{-1}}$. The intrinsic velocity dispersions, $\sigma_1$ and $\sigma_2$, increase with $|Z|$ and are
modeled as $a+b|Z|^c$, with the best-fit parameters $a$, $b$ and $c$ listed in Table 1 from \cite{2010ApJ...716....1B} ($c=2.0$ for both).

The mean values for the $v_R$ and $v_Z$ components were found to be zero, as expected. Their intrinsic dispersions for disk
stars vary with $Z$ and are described by the following expressions:
\begin{equation}
    \sigma_{R} = 40 + 5.0 \,\left|\frac{Z}{\rm kpc}\right|^{1.5} \,\,\, {\rm km~s^{-1}}.
\label{eq:DdispR}
\end{equation}
and  
\begin{equation}
    \sigma_{Z} = 25 + 4.0 \,\left|\frac{Z}{\rm kpc}\right|^{1.5} \,\,\, {\rm km~s^{-1}}.
\label{DdispZ}
\end{equation} 

They also found that the $\sigma_{R}/\sigma_{\phi}$ ratio for disk stars has a constant value of $\sim1.35$ for $Z<1.5$~kpc,
and decreases steadily at larger $Z$ to about $1$ at $Z\sim4$~kpc.

We note that for stars in the analyzed FGKM sample, the mean age increases and the mean metallicity decreases with
the distance from the Galactic plane, with the latter also observed in analyzed SDSS data. For attempts to interpret these
correlations using a radial migration model for disk stars, see \cite{2011ApJ...737....8L}.

In the next section, we compare these models to \textit{Gaia} measurements for stars in the FGKM and red giant samples.
In particular, we aim to test whether the above fits to the median rotational velocity and velocity dispersion,
constrained by luminous blue stars with $r>14$ used in the SDSS study, can be successfully extrapolated closer
to the Galactic plane, as probed by \textit{Gaia} measurements for stars with $r<14$ in the FGKM sample.

\subsubsection{Kinematic model for halo stars\label{sec:BondModelsHalo}}

Bond et al. found that the kinematics of halo stars in their sample, which was confined to galacto-centric radial distances in the range
5 -- 20 kpc, admit an exceedingly simple model description. Their dataset was fully explained with no net rotation
and a triaxial velocity ellipsoid model that is invariant in spherical coordinates, with the following velocity dispersions:
$\sigma_r= 141$~km~s$^{-1}$, $\sigma_\phi= 85$ km~s$^{-1}$, and $\sigma_\theta= 75$ km~s$^{-1}$,
with uncertainties of about 5 -- 10 km~s$^{-1}$. This remarkable alignment of the halo velocity ellipsoid with spherical
coordinates (halo stars “know” where the Galactic center is)  represents a strong constraint on the
shape of gravitational potential—the potential must be close to spherically symmetric at least within about 20 kpc
from the Galactic Center (for more details and references, see
\citealt{2014ApJ...794..151L, 2019MNRAS.489..910E}).

\section{Results\label{sec:results}}

Here we test the validity of the Bond et al. kinematic models for disk and halo stars, as enabled by the high-metallicity FGKM sample
and the red giant sample introduced in the preceding section.
We first discuss the behavior of the first two statistical moments (mean and dispersion) for 3-dimensional velocity distribution
and then compare the predicted and observed shapes of the rotational velocity distribution in more detail. We test
model predictions for the behavior of quantities directly measured by \textit{Gaia}, stellar radial velocities and proper motions, across
the whole sky. Finally, we verify that the behavior of \textit{Gaia} data for halo stars is consistent with a velocity ellipsoid that is
spatially invariant when expressed in spherical coordinates.

\subsection{Comparison of predicted and observed 3-dimensional velocity distribution for disk stars}

\begin{figure*}[ht]
\includegraphics[width=0.999\linewidth,angle=0]{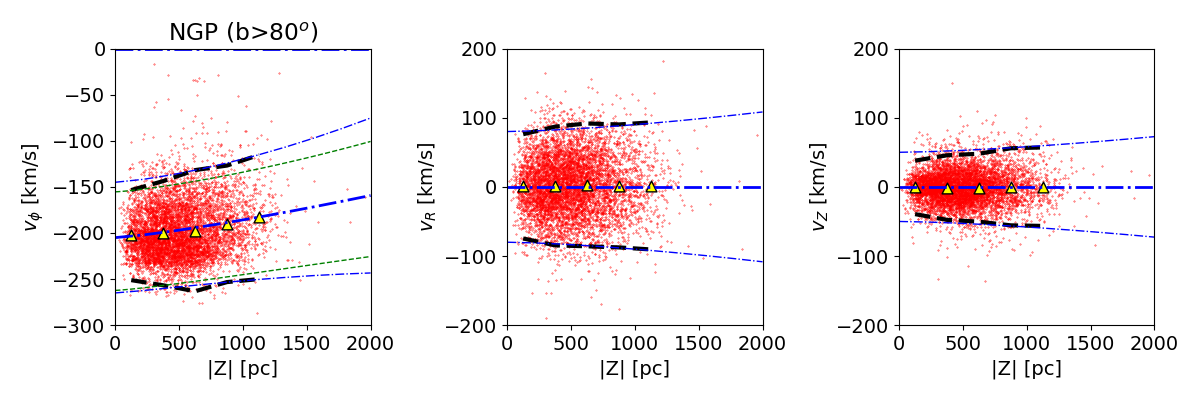}
\includegraphics[width=0.999\linewidth,angle=0]{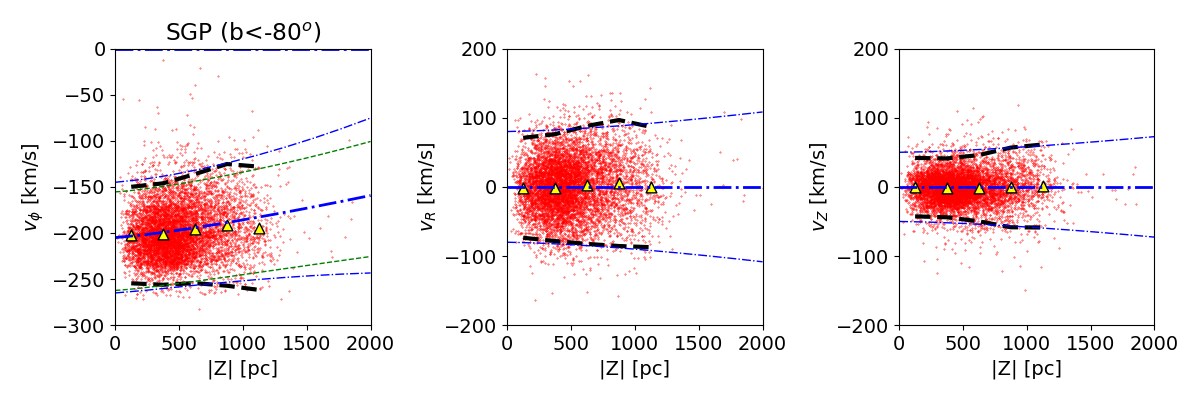}
\caption{Variation of cylindrical velocity components with distance from the plane, $|Z|$, for FGKM stars with measured radial velocities (red dots). 
The top row shows 6,548 stars towards the North Galactic Pole ($b >80^\circ$) and the bottom row shows 7,578 stars towards the South Galactic Pole
($b < -80^\circ$). Triangles show the mean values in bins of $|Z|$ and the thick dashed lines show the $\pm2\sigma$ envelope around means, where
$\sigma$ is the standard deviation for each bin (\ie, velocity dispersion). The thick dot-dashed lines are models for the mean velocity (0 for $v_R$ and
$v_Z$, and given by eq.~\ref{eq:Dlag} for $v_\phi$). The thin dot-dashed lines show the $\pm2\sigma$ envelope, with $\sigma$ (velocity dispersion)
given by eqs.~\ref{eq:Ddisp}, \ref{eq:DdispR}, and \ref{DdispZ}, for $\phi$, $R$ and $Z$ panels, respectively.  
The thin dashed lines in the two left panels
also show $\pm2\sigma$ envelopes, but with the velocity dispersion computed using eq.~\ref{eq:pDvPhi} (see Sect. ~\ref{sec:vshape}).} 
\label{fig:3v} 
\end{figure*}

We first emulate the top left panels in Figures 5, 7 and 11 from the Bond et al. study. We select stars towards the North and South
Galactic poles ($|b|>80^\circ$) and show the variation of their velocity components with $|Z|$ in \autoref{fig:3v}. Due to the
galactic latitude constraint, the $v_Z$ component is dominated by the radial velocity measurements while the other two components
are dominated by proper motion measurements. Compared to the Bond et al. high-luminosity sample, the distance range probed
by \textit{Gaia's} FGKM sample is closer: from about 50 pc to about 1 kpc vs. 0.8--5 kpc (at the far end, the SDSS sample is limited by
sample contamination due to halo stars).

The Bond et al. model predictions for the velocity mean and dispersion are in good agreement with \textit{Gaia's} observations. In particular, the
gradient of the rotational velocity with $|Z|$ is evident, and the extrapolation of SDSS rotational velocity measurements to $Z=0$
is quantitatively supported by \textit{Gaia}. The same conclusions remain valid when the full sample is considered (that is, without the $|b|>80^\circ$ restriction). 
% {\bf 
The level of discrepancies between the model and \textit{Gaia} measurements seen in \autoref{fig:3v}, of the order 10\%, is 
comparable to the level of north vs. south asymmetries found recently in other \textit{Gaia}-based studies 
\citep{2020A&A...643A..75S, 2022MNRAS.511.3863E, 2023MNRAS.520.3329L}.
% }

\begin{figure*}[ht]
\includegraphics[width=0.999\linewidth,angle=0]{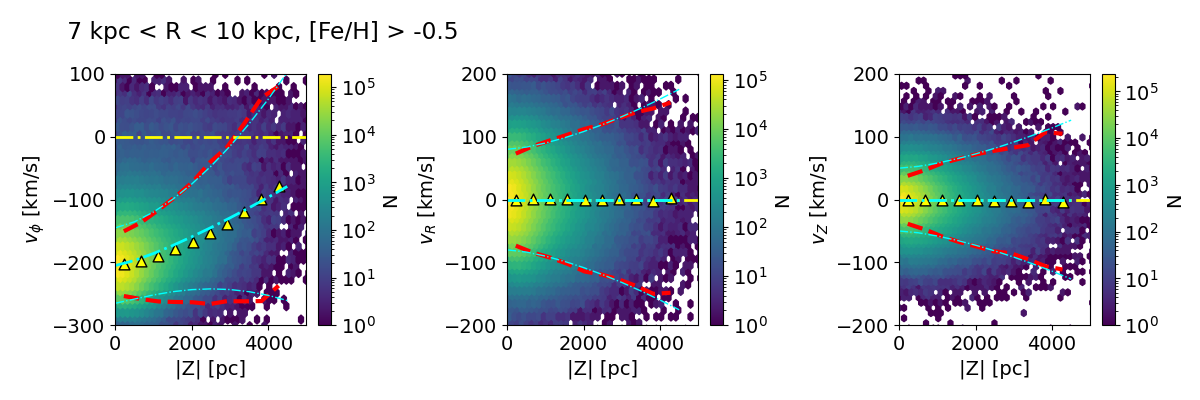} 
\caption{Similar to \autoref{fig:3v}, except that here about five times larger distances are probed with 4.3 million red giant stars
  that have [Fe/H]$>-0.5$ and cylindrical galacto-centric radius between 7 kpc and 10 kpc. The meaning of symbols and lines
  is the same as in \autoref{fig:3v}, and exactly the same models are shown.} 
\label{fig:3vRGs} 
\end{figure*}

The $<$1 kpc distance range probed by FGKM stars is significantly extended with the red giant sample, as illustrated in
\autoref{fig:3vRGs}. In conclusion, the Bond et al. models for the velocity mean and dispersion of disk stars appear validated
in the $|Z|$ range 0--5 kpc, without any appreciable north-south asymmetry.

\subsection{Comparison of predicted and observed shapes of the rotational velocity distribution for disk stars\label{sec:vshape}}

\begin{figure}
\centering
\includegraphics[width=\linewidth,angle=0]{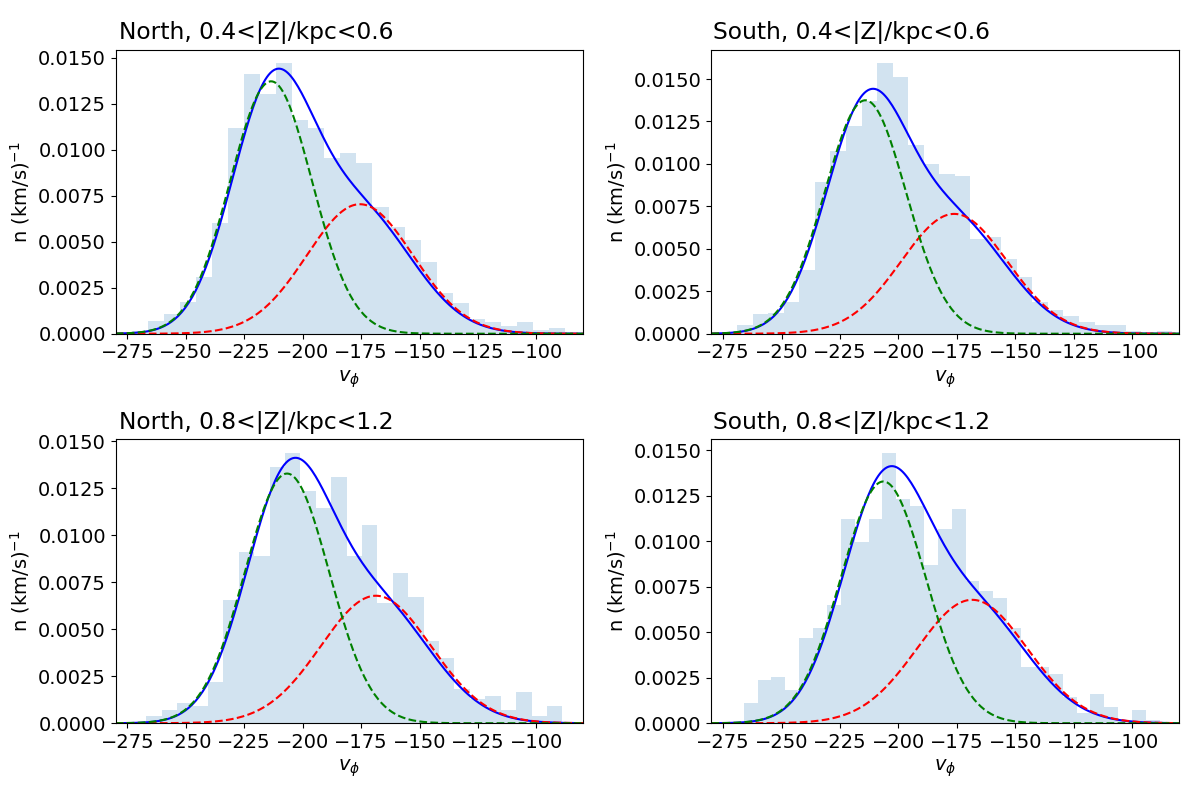}
\caption{Comparison of the observed rotational velocity distribution ($v_\phi$) for FGKM stars, shown as histograms, and a two-component model (dashed
lines: individual components; solid line: their sum) given by \autoref{eq:pDvPhi} (using updated parameters discussed in text; the same
$Z$-dependent model is shown in all four panels). The four panels show results for two $|Z|$ bins (top row vs. bottom row) and towards North and South
($|b|>80^\circ$, left vs. right). The sample sizes are about 4,000 stars for the nearer $|Z|$ bin and about 900 stars for the more distant bin.} 
\label{fig:vPhiShape} 
\end{figure}

As already suggested in \cite{2008ApJ...684..287I} and confirmed by Bond et al. (see their Figure 6), the shape of the rotational velocity
distribution for disk stars is strongly non-Gaussian (skewed) and it evolves with the distance from the Galactic plane. Histograms based
on \textit{Gaia's} data shown in \autoref{fig:vPhiShape} confirm this expectation. Furthermore, new data still admit detailed quantitative
modeling of observed distributions as a sum of two Gaussians, with a fixed normalization ratio and a fixed offset of their mean values (see
\autoref{eq:pDvPhi}). Since \textit{Gaia's} velocity measurement errors are negligible in this context (compared to the much larger width of the observed
velocity distributions; for validation using quasar data, see Appendix), we have reoptimized fits by allowing six parameters to vary. Their best-fit values remain close to the SDSS
values: $f_1=0.40$, $f_2=0.60$, $v_0 = -186 {\rm \, km~s^{-1}}$, $\Delta v_n = 38 {\rm \, km~s^{-1}}$, $a_1 = 22 {\rm \, km~s^{-1}}$ and $a_2= 17
{\rm \, km~s^{-1}}$.  

When the resulting re-optimized pdf, $p(x=v_\phi|Z)$, is used to evaluate the mean $\langle v_\phi \rangle$ and its dispersion
$\sigma_{\phi}$  as functions of $Z$, there is no discernible change for the former compared to \autoref{eq:Dlag}, while the latter is
about 20\% smaller than values given by \autoref{eq:Ddisp} (for illustration of the difference, see the two left panels in \autoref{fig:3v}).
Since there is no implied physics in this two-component statistical model of the skewed velocity distribution, minor numerical adjustments
of the few free model parameters are probably inconsequential. \\

\begin{figure}
\centering
\includegraphics[width=\linewidth,angle=0]{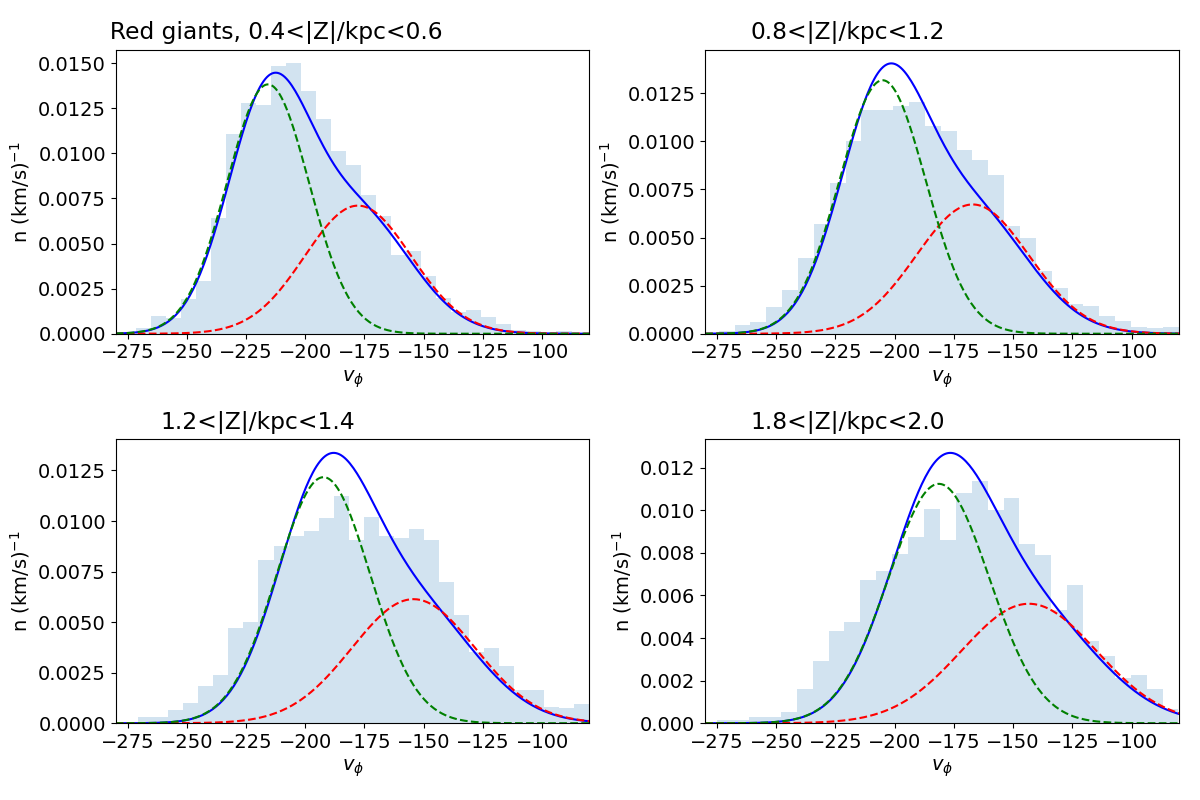}
\caption{Analogous to \autoref{fig:vPhiShape}, except for a subsample of the red giant sample shown in \autoref{fig:3vRGs} restricted to $|b|>60^\circ$.
  The top two panels show the same $|Z|$ bins as in \autoref{fig:vPhiShape}, while the two bottom panels extend to larger $|Z|$. 
  The sample sizes vary from 10,000 stars to 2,000 stars for the most distant bin.} 
\label{fig:vPhiShapeRG} 
\end{figure}

These conclusions based on the FGKM sample are supported by the behavior of the red giant sample, as illustrated in \autoref{fig:vPhiShapeRG}. 
In particular, the vertical gradient of rotational velocity is evident. 

\subsection{Comparison of predicted and observed radial velocity and proper motion sky distributions for disk stars\label{sec:allsky}}

\begin{figure}[!ht]
\centering
\includegraphics[width=0.49\linewidth,angle=0]{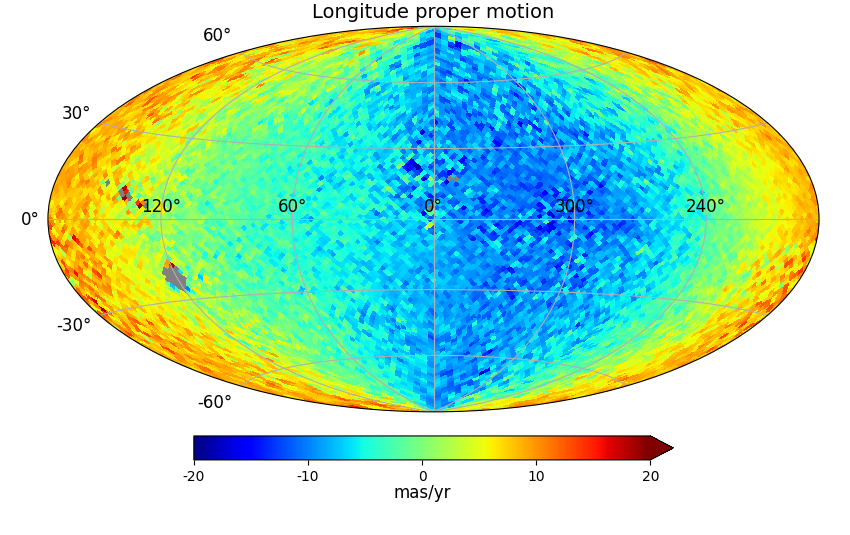}
\includegraphics[width=0.49\linewidth,angle=0]{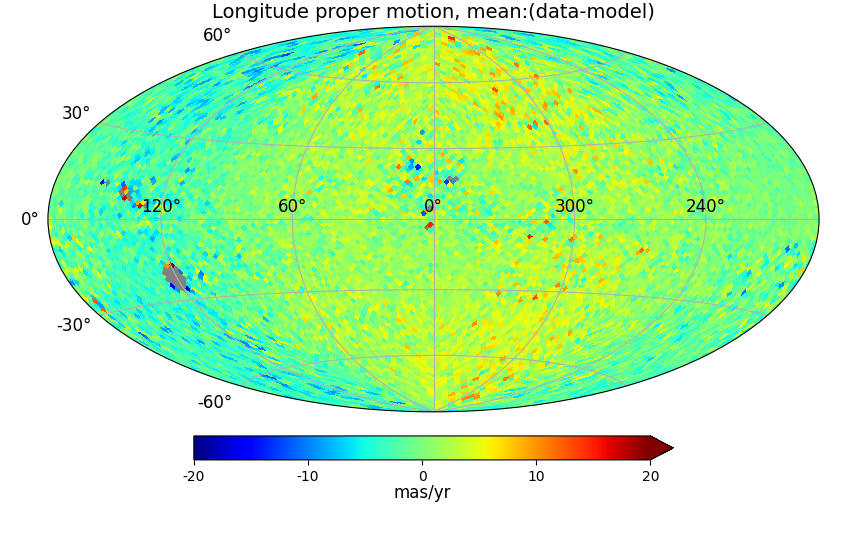}
\includegraphics[width=0.49\linewidth,angle=0]{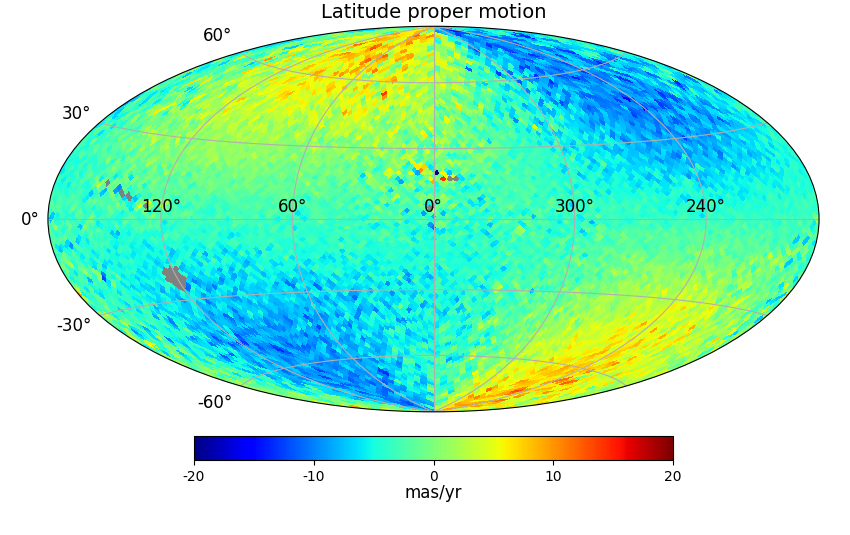}
\includegraphics[width=0.49\linewidth,angle=0]{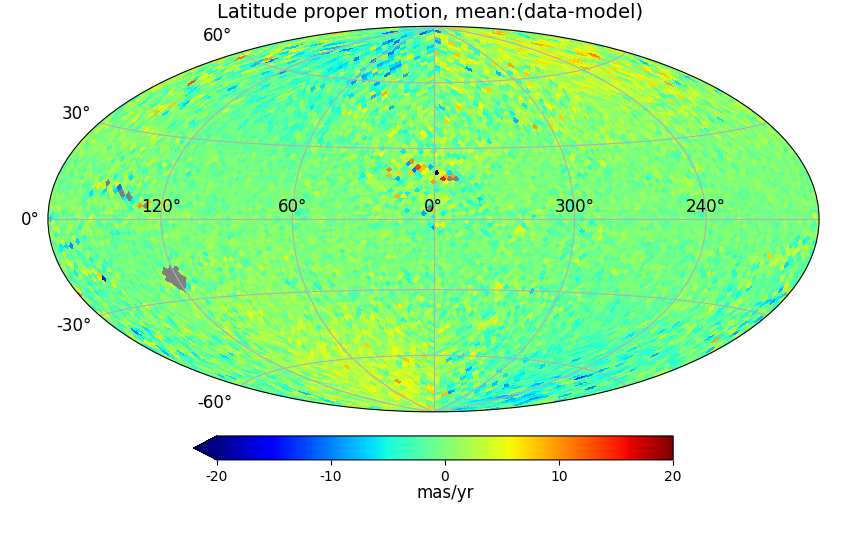}
\includegraphics[width=0.49\linewidth,angle=0]{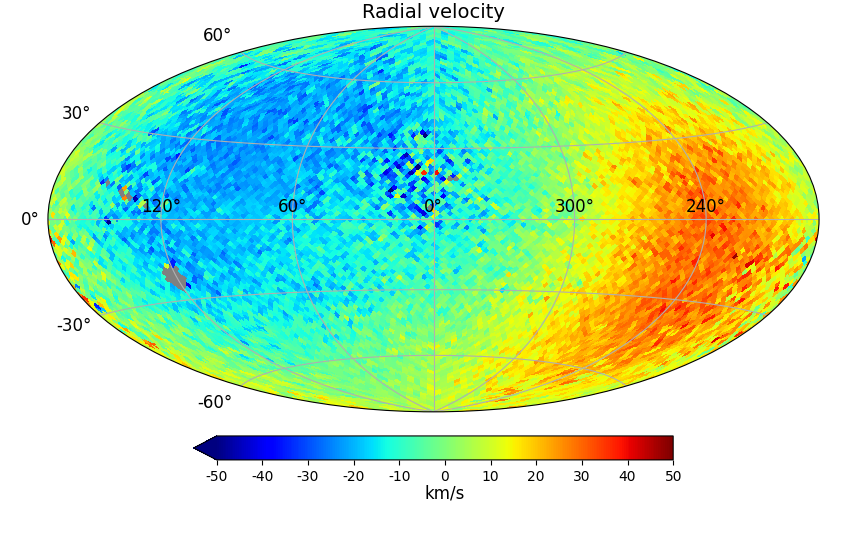}
\includegraphics[width=0.49\linewidth,angle=0]{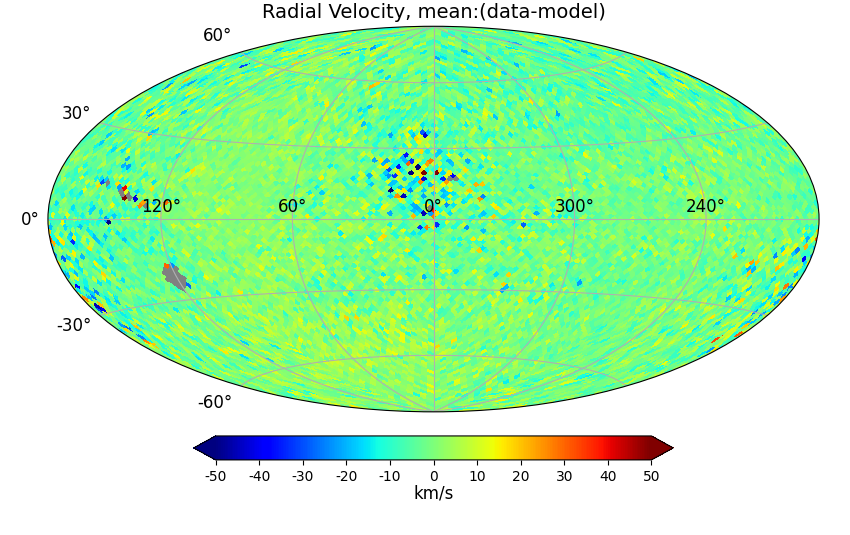}
\caption{Observed distribution of mean proper motion and mean radial velocity per pixel. The top two left panels show observed distributions of mean proper motion per pixel (top: longitude; middle: latitude)
for a subsample of 400,782 stars with good measurements in the 0.4 -- 0.6~kpc distance bin. The bottom left panel shows the observed distribution of mean radial velocity. The HEALPix nside=32 maps are shown in the Hammer
projection of galactic coordinates. The maps on the right show residuals (using the same color-coding scale) after
subtracting model values computed using eq.~\ref{eq:Dlag} for $v_\phi$ and with $\langle v_R \rangle = \langle v_Z \rangle = 0$.} 
\label{fig:skyComparison} 
\end{figure}

As the final test, we extrapolate the Bond et al. model from the quarter of the sky observed by SDSS and compare its predictions for
the variation of mean proper motions and radial velocities to \textit{Gaia's} measurements across the whole sky. The three left panels in
\autoref{fig:skyComparison} show mean proper motion components per pixel and mean radial velocity per pixel for stars with
distances in the range of 400 pc to 600 pc. Their strong variation across the sky is evident and is mostly due to projection effects of the
solar motion (and a little bit due to spatial variation of the rotational velocity component with $|Z|$). The panels on the right
show residuals after subtracting proper motion and radial velocity values predicted by the Bond et al. model described in Sect. ~\ref{sec:BondModels}.
As evident, the model fully captures the observed behavior. The only potentially significant feature is observed for radial velocity residuals towards the Galactic
center, at about $0 < b < 30^\circ$. This region was not observed by SDSS and we are not aware of any kinematic or other stellar features
reported for that region (\eg, the famous Sgr dwarf galaxy is located at negative galactic latitudes). A possibility that these residuals
reflect faulty measurements could be, at least in principle, easily checked because most of these stars are relatively bright ($r\sim14$). 

We note that the data maps for stars with distances in the range of 800 pc to 1.2 kpc look qualitatively the same, except that the magnitude
of proper motion is smaller by about a factor of two, as expected. The maps of residuals appear the same, except for the radial velocity
residual ``feature'', which is observed at about 15 degrees higher latitudes than for the first distance bin.

\begin{figure}[ht]
\includegraphics[width=0.49\linewidth,angle=0]{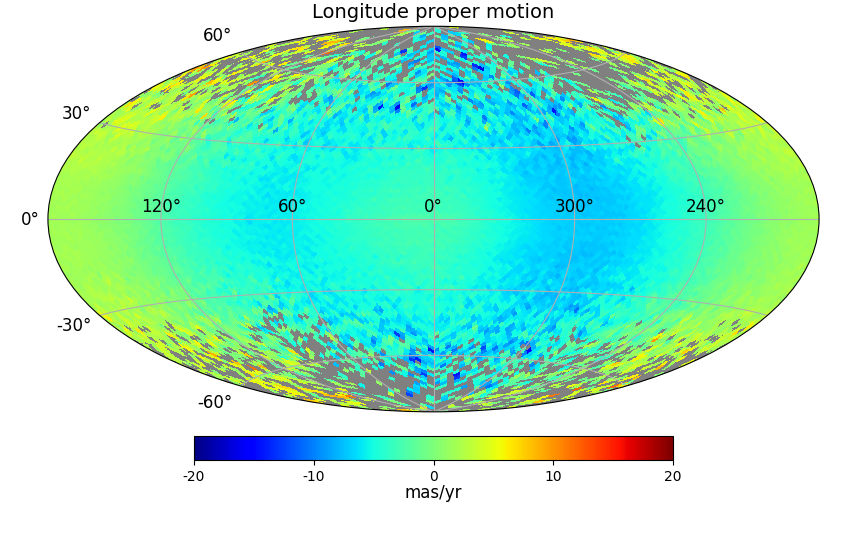}
\includegraphics[width=0.49\linewidth,angle=0]{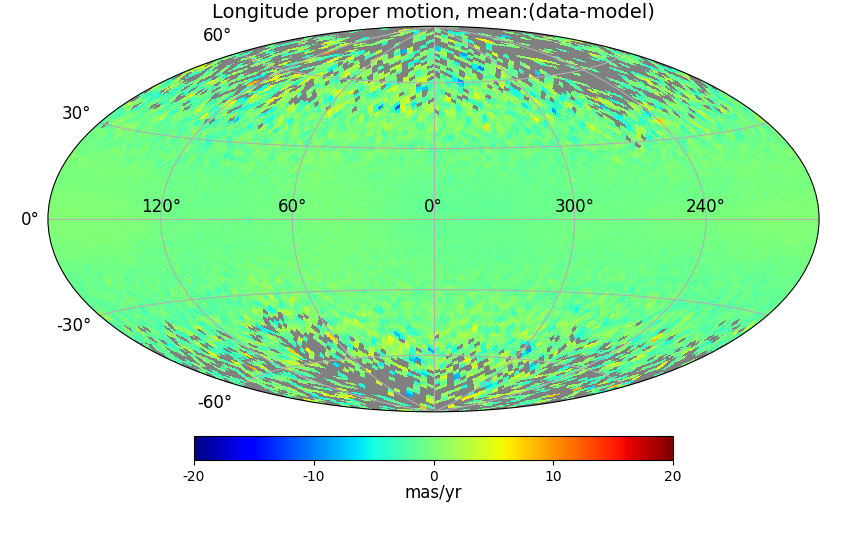}
\includegraphics[width=0.49\linewidth,angle=0]{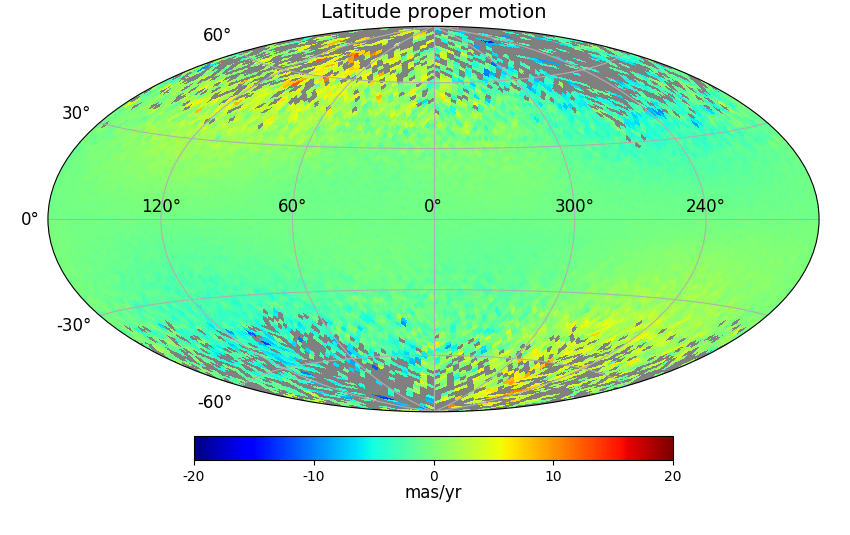}
\includegraphics[width=0.49\linewidth,angle=0]{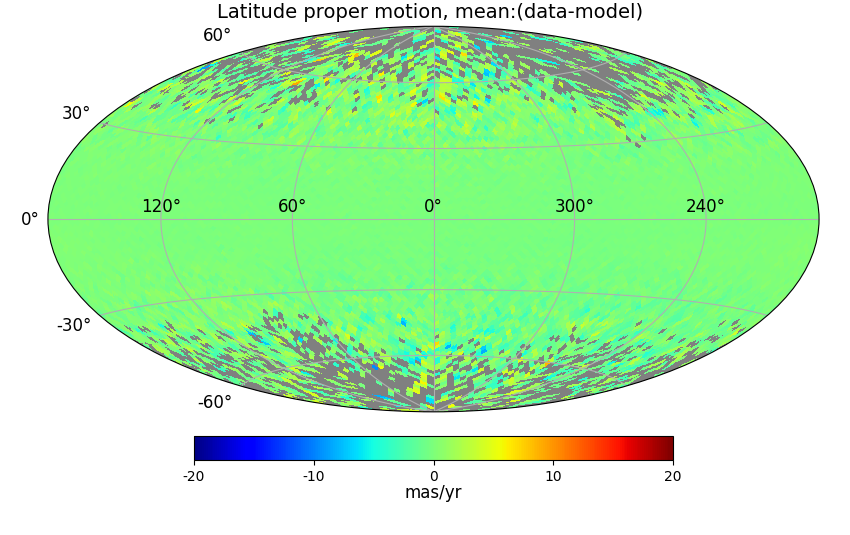}
\includegraphics[width=0.49\linewidth,angle=0]{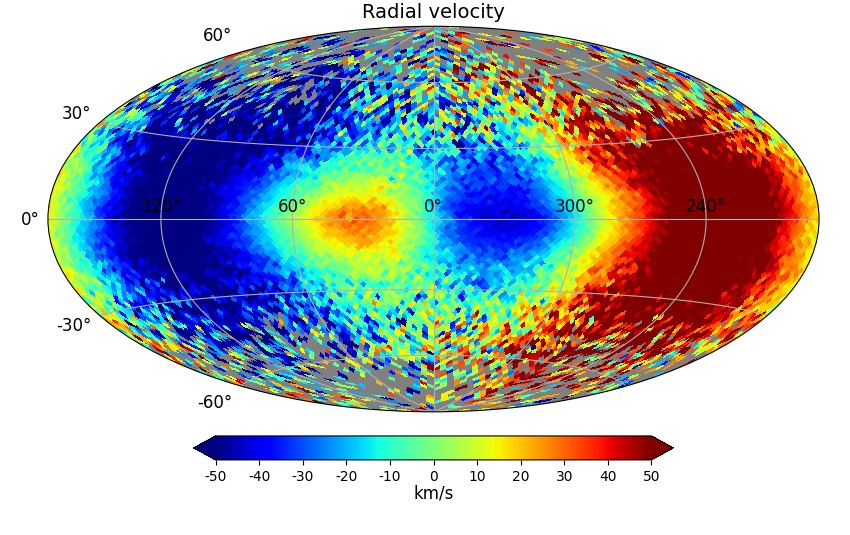}
\includegraphics[width=0.49\linewidth,angle=0]{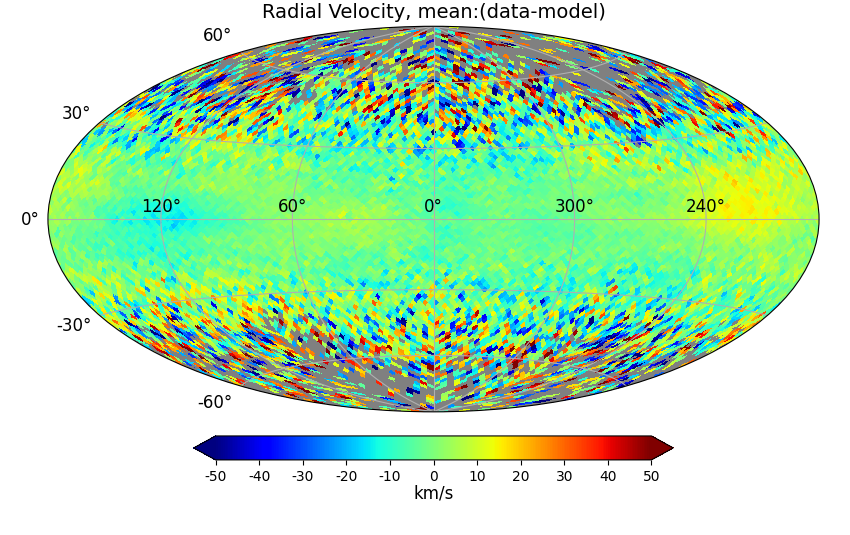}
\caption{Analogous to \autoref{fig:skyComparison}, except for a sample of 1.05 million red giants with [Fe/H]$>-0.5$
  in the 2.8 -- 3.2 kpc distance bin. Note that observed proper motions are much smaller due to about six times larger distances,
  while radial velocity variation with galactic longitude is much stronger due to the projection effects of the rotational velocity component.} 
\label{fig:skyComparisonRG} 
\end{figure}

The red giant sample can be used to extend this comparison to distances where the rotational velocity component is much
smaller than locally. We first established that the behavior for red giants in the 0.4 -- 0.6 kpc distance bin is essentially identical
as shown for the FGKM sample in \autoref{fig:skyComparison}. The corresponding plot for red giants in the 2.8 -- 3.2 kpc distance bin 
is shown in \autoref{fig:skyComparisonRG}. Due to about six times larger distances,  the morphology of data panels is rather different
from that in \autoref{fig:skyComparison} (especially for radial velocity, where projection effects of rotational velocity component
dominate). Nevertheless, the model for disk kinematics described in
Sect. ~\ref{sec:BondModels} fully captures the data behavior, as shown by vanishing maps of (data-model) residuals.

\subsection{Comparison of predicted and observed 3-dimensional velocity distribution for halo stars}

\begin{figure*}[!ht]
    \centering
    \includegraphics[width=0.999\linewidth,angle=0]{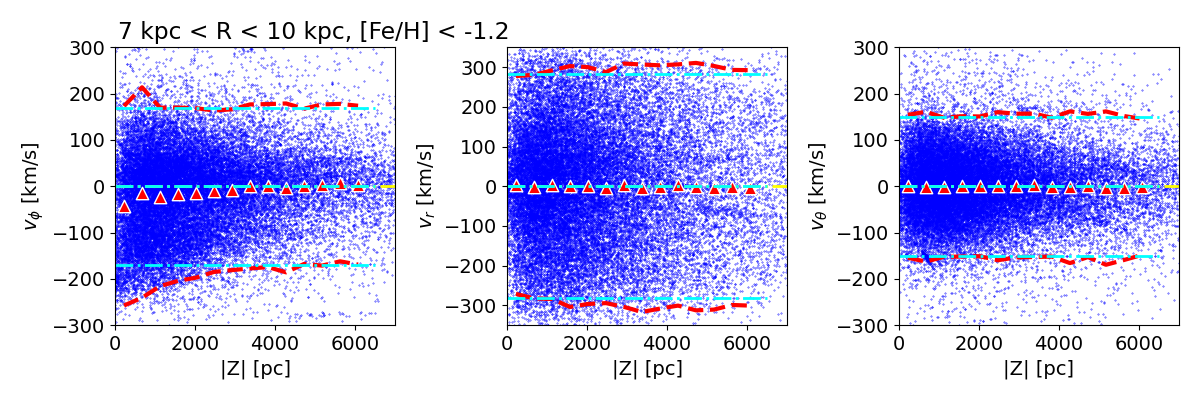}
    \caption{Variation of the spherical velocity components with distance from the plane, $|Z|$, for $\sim$34,000 candidate halo red giants from the solar cylinder with measured radial velocities and [Fe/H]$<-1.2$. Triangles show the mean values in bins of $|Z|$ and the thick dashed lines show the $\pm2\sigma$ envelope around means, where $\sigma$ is the standard deviation for each bin (\ie, velocity dispersion). The dot-dashed lines represent the halo velocity ellipsoid model described in Section~\ref{sec:BondModelsHalo}. \label{fig:3vRGhalo}}
\end{figure*}

\begin{figure}[!h]
\centering
\includegraphics[width=0.99\linewidth,angle=0]{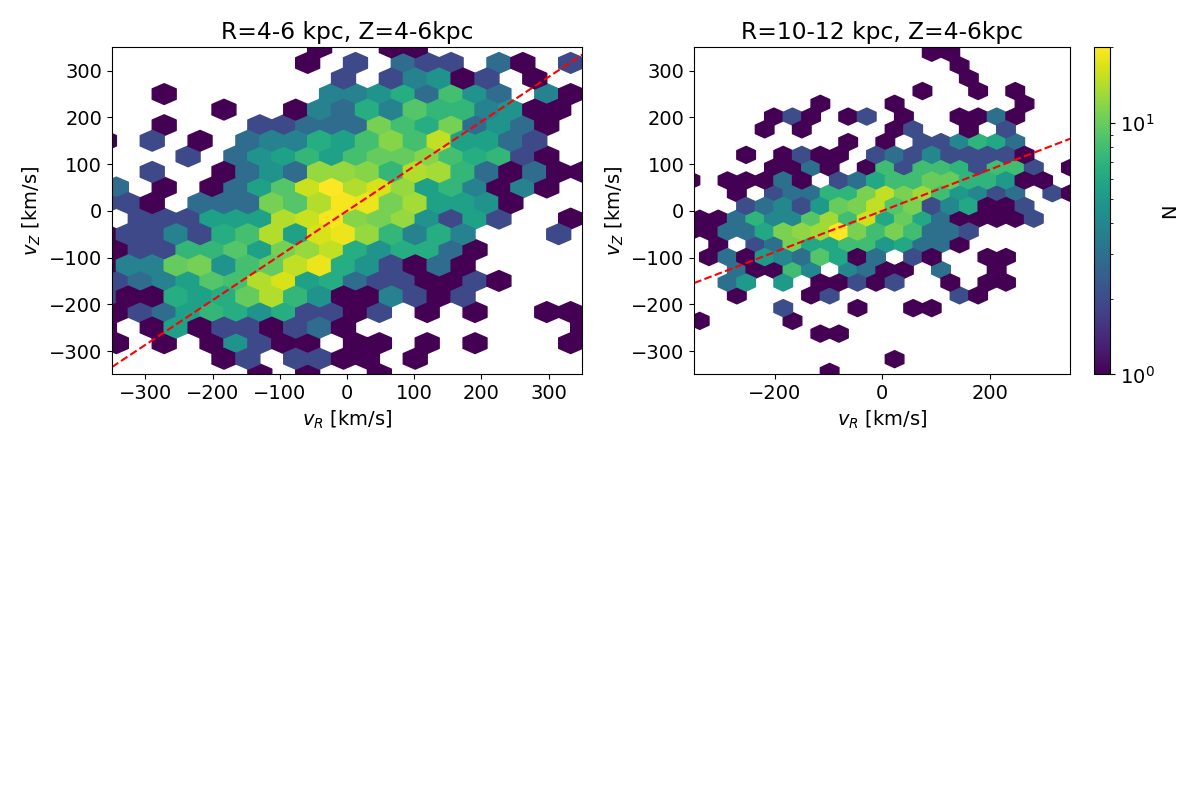}
\vskip -1in   
\includegraphics[width=0.99\linewidth,angle=0]{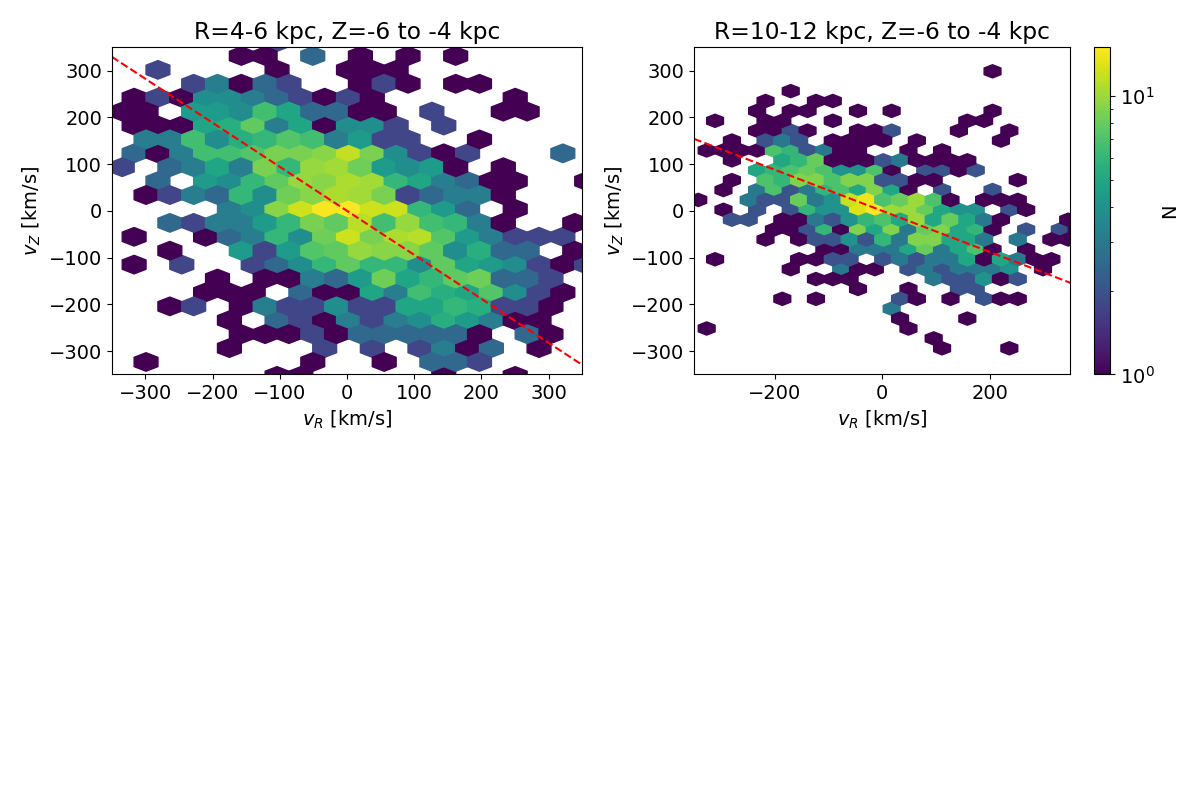}
\vskip -1in   
\caption{Illustration of the change of orientation of velocity ellipsoid in cylindrical coordinates for halo stars,
  selected to have [Fe/H]$<-1.2$ and selected from a narrow range of cylindrical coordinates. The number of stars is about 1,600 for bins with smaller $R$ (left) and 800 for bins on the right.
  The top row shows bins above the plane and the bottom row their symmetric counterparts below the plane
  (the Sun's position is in the middle of the figure). 
  The dashed lines mark the direction towards the Galactic center. The change in the velocity ellipsoid tilt
  in cylindrical coordinates is evident; however, the velocity ellipsoid in spherical coordinates is invariant
  throughout the probed volume, as illustrated by these four spatial bins.} 
\label{fig:haloTilt} 
\end{figure}

Bond et al. found that the kinematics of halo stars can be modeled with a triaxial velocity ellipsoid that is invariant in
spherical coordinates. Motivated by this finding, we show in
\autoref{fig:3vRGhalo} three spherical velocity components for halo red giants as functions of distance from the Galactic plane. Unlike the strong dependence on the velocity dispersion
on $|Z|$ for disk stars seen in Figures~\ref{fig:3v} and \ref{fig:3vRGs}, velocity dispersion for halo stars as measured by
\textit{Gaia} is spatially invariant, confirming earlier results by Bond et al. We note that small deviation of mean rotational
velocity from zero at $|Z|<1$ kpc seen in the left panel is probably due to sample contamination by much more numerous
disk stars (at $|Z|=0$, halo stars contribute only about 0.5\% of the total count, see Table 10 in \citealt{2008ApJ...673..864J}).  

\textit{Gaia} data analyzed here provide strong support for a halo velocity ellipsoid that is invariant in spherical coordinates.
When the velocity ellipsoid is expressed in cylindrical coordinates instead, a strong covariance is seen between the $v_Z$ and $v_R$
components. We illustrate this covariance in
\autoref{fig:haloTilt}.
The observed tilt varies with position such that the
velocity ellipsoid points towards the Galactic center (see also \citealt{2019MNRAS.489..910E}). 

\section{Discussion and conclusions\label{sec:disc}}
\begin{figure*}[!ht]
\includegraphics[width=0.999\linewidth,angle=0]{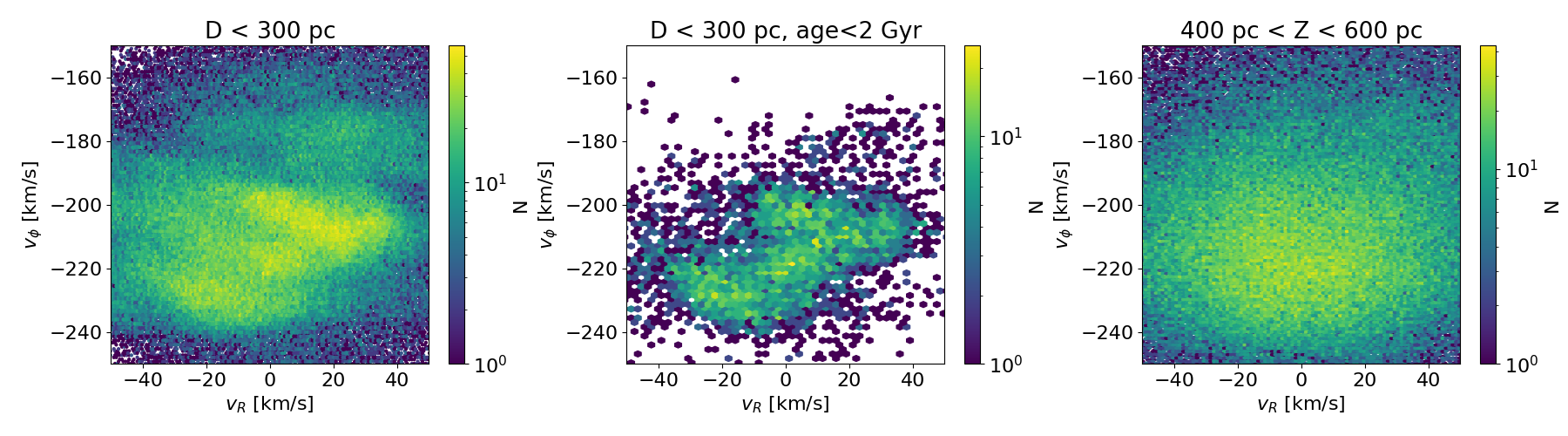} 
\caption{A comparison of distributions in the $v_\phi$ vs. $v_R$ diagram for three different samples of disk stars: a nearby sample
of 276,453 stars at distances below 300 pc (left), its subsample consisting of 6,108 stars younger than 2 Gyr (middle), and a
sample of 237,660 stars at distances from the Galactic plane in the range of 400--600 pc. Complex structure, known in the literature
as ``Eggen's moving groups'', is evident at distances below 300 pc. 
For example, the feature at ($v_R=-25, v_\phi=-230$) corresponds to the Sirius moving group, the feature at ($v_R=0, v_\phi=-220$)
is the Coma B moving group, the feature at ($v_R=0, v_\phi=-200$) is the Pleiades moving group, and the Hyades moving group
is seen at ($v_R=30, v_\phi=-210$).} 
\label{fig:eggen} 
\end{figure*}

We validated the \citet{2010ApJ...716....1B} kinematic models for the Milky Way's disk and halo stars with \textit{Gaia} Data Release 3 data. For disk stars, the gradient of rotational velocity with $|Z|$ is evident in \textit{Gaia's} data, and the extrapolation of SDSS rotational velocity
measurements to $Z=0$ is also supported by \textit{Gaia}. The models for the velocity mean and dispersion are validated in the
$|Z|$ range of 0 -- 5 kpc, and show no appreciable north-south asymmetry.
% {\bf 
However, we cannot 
strongly exclude north vs. south symmetries at the level of about 10\%, as recent 
\textit{Gaia}-based studies suggest\citep{2020A&A...643A..75S, 2022MNRAS.511.3863E, 2023MNRAS.520.3329L}.
% }

The $v_\phi$ vs. $v_R$ distribution near the Galactic plane (closer than a few hundred parsecs) is very complex
and cannot be described by a standard Schwarzschild ellipsoid (for details and references, see Sect. 3.1 in \citealt{2008ApJ...684..287I}). 
The panels in
\autoref{fig:eggen} are analogous to Figure 4 from \citet{2010ApJ...716....1B}. The so-called ``Eggen's moving groups''
\citep{1996AJ....112.1595E} are clearly visible at distances less than 300 pc and cannot be described in detail by the models
discussed here. 

The all-sky proper motion and radial velocity test, based on the FGKM sample discussed in Sect. \ref{sec:allsky}, is encouraging. However, we caution that
most observed variations of measured quantities across the sky result from projection effects of solar motion, with the spatial variation of the rotational velocity component with $|Z|$ contributing minimally. Therefore, this test mainly supports our adopted solar motion values 
taken from \cite{1998MNRAS.298..387D}, derived using Hipparcos data. The test based on red giants is stronger, extending to about six times larger distances.

For halo stars, the spatial invariance of their velocity ellipsoid (when expressed in spherical coordinates) is also confirmed by \textit{Gaia}
data at galacto-centric radial distances of up to 15 kpc. For related work, please see \cite{2019MNRAS.489..910E}. 

Given these successful tests, the Bond et al. kinematic models for disk and halo stars are adequate for implementation
in simulated catalogs of the Milky Way's stellar content, such as the recent TRILEGAL-based simulated LSST catalog by
\cite{2022ApJS..262...22D}. As discussed by \cite{2012ARA&A..50..251I} and \cite{2019ApJ...873..111I},
LSST will provide its own kinematic constraints, with numerous main-sequence stars out to distances of about 30 kpc,
significantly farther than possible with \textit{Gaia} data due to several magnitudes fainter survey limit.

\begin{acknowledgements}
We are thankful to the 2023 Vatican Observatory Summer School in Observational Astronomy and Astrophysics, where this work originated. 
B.D. acknowledges funding provided by the Universidad de la República through the CSIC's Mobility and Academic Exchange Program.
S.C. acknowledges funding provided by the University of Delaware through the Graduate Student Travel Award.
\v{Z}.I. acknowledges funding by the Fulbright Foundation and thanks the Rudjer Bo\v{s}kovi\'{c} Institute for hospitality. We thank Dr. Lovro Palaversa for communicating his early results on the comparison of SDSS and \textit{Gaia} distance scales. 

Funding for the SDSS and SDSS-II has been provided by the Alfred
P. Sloan Foundation, the Participating Institutions, the National
Science Foundation, the U.S. Department of Energy, the National
Aeronautics and Space Administration, the Japanese Monbukagakusho, the
Max Planck Society, and the Higher Education Funding Council for
England. The SDSS Web Site is \url{https://www.sdss.org/}.

This work has made use of data from the European Space Agency (ESA)
mission \textit{Gaia} \url{https://www.cosmos.esa.int/gaia}, processed by the \textit{Gaia}
Data Processing and Analysis Consortium (DPAC,
\url{https://www.cosmos.esa.int/web/gaia/ dpac/consortium}). Funding for the
DPAC has been provided by national institutions, in particular, the
institutions participating in the \textit{Gaia Multilateral Agreement}.
\end{acknowledgements}


\bibliographystyle{aa}
\bibliography{paper.bib}

\begin{thebibliography}{19}
\expandafter\ifx\csname natexlab\endcsname\relax\def\natexlab#1{#1}\fi

\bibitem[{{Andrae} {et~al.}(2023){Andrae}, {Rix}, \&
  {Chandra}}]{2023ApJS..267....8A}
{Andrae}, R., {Rix}, H.-W., \& {Chandra}, V. 2023, \apjs, 267, 8

\bibitem[{{Bailer-Jones} {et~al.}(2021){Bailer-Jones}, {Rybizki}, {Fouesneau},
  {Demleitner}, \& {Andrae}}]{2021AJ....161..147B}
{Bailer-Jones}, C.~A.~L., {Rybizki}, J., {Fouesneau}, M., {Demleitner}, M., \&
  {Andrae}, R. 2021, \aj, 161, 147

\bibitem[{{Bond} {et~al.}(2010){Bond}, {Ivezi{\'c}}, {Sesar}, {Juri{\'c}},
  {Munn}, {Kowalski}, {Loebman}, {Ro{\v{s}}kar}, {Beers}, {Dalcanton},
  {Rockosi}, {Yanny}, {Newberg}, {Allende Prieto}, {Wilhelm}, {Lee},
  {Sivarani}, {Majewski}, {Norris}, {Bailer-Jones}, {Re Fiorentin}, {Schlegel},
  {Uomoto}, {Lupton}, {Knapp}, {Gunn}, {Covey}, {Allyn Smith}, {Miknaitis},
  {Doi}, {Tanaka}, {Fukugita}, {Kent}, {Finkbeiner}, {Quinn}, {Hawley},
  {Anderson}, {Kiuchi}, {Chen}, {Bushong}, {Sohi}, {Haggard}, {Kimball},
  {McGurk}, {Barentine}, {Brewington}, {Harvanek}, {Kleinman}, {Krzesinski},
  {Long}, {Nitta}, {Snedden}, {Lee}, {Pier}, {Harris}, {Brinkmann}, \&
  {Schneider}}]{2010ApJ...716....1B}
{Bond}, N.~A., {Ivezi{\'c}}, {\v{Z}}., {Sesar}, B., {et~al.} 2010, \apj, 716, 1

\bibitem[{{Dal Tio} {et~al.}(2022){Dal Tio}, {Pastorelli}, {Mazzi},
  {Trabucchi}, {Costa}, {Jacques}, {Pieres}, {Girardi}, {Chen}, {Olsen},
  {Juric}, {Ivezi{\'c}}, {Yoachim}, {Clarkson}, {Marigo}, {Rodrigues},
  {Zaggia}, {Barbieri}, {Momany}, {Bressan}, {Nikutta}, \& {da
  Costa}}]{2022ApJS..262...22D}
{Dal Tio}, P., {Pastorelli}, G., {Mazzi}, A., {et~al.} 2022, \apjs, 262, 22

\bibitem[{{Dehnen} \& {Binney}(1998)}]{1998MNRAS.298..387D}
{Dehnen}, W. \& {Binney}, J.~J. 1998, \mnras, 298, 387

\bibitem[{{Eggen}(1996)}]{1996AJ....112.1595E}
{Eggen}, O.~J. 1996, \aj, 112, 1595

\bibitem[{{Everall} {et~al.}(2022){Everall}, {Belokurov}, {Evans}, {Boubert},
  \& {Grand}}]{2022MNRAS.511.3863E}
{Everall}, A., {Belokurov}, V., {Evans}, N.~W., {Boubert}, D., \& {Grand}, R.
  J.~J. 2022, \mnras, 511, 3863

\bibitem[{{Everall} {et~al.}(2019){Everall}, {Evans}, {Belokurov}, \&
  {Sch{\"o}nrich}}]{2019MNRAS.489..910E}
{Everall}, A., {Evans}, N.~W., {Belokurov}, V., \& {Sch{\"o}nrich}, R. 2019,
  \mnras, 489, 910

\bibitem[{{Gaia Collaboration} {et~al.}(2021){Gaia Collaboration}, {Brown},
  {Vallenari}, {Prusti}, {de Bruijne}, {Babusiaux}, {Biermann}, {Creevey},
  {Evans}, {Eyer}, {Hutton}, {Jansen}, {Jordi}, {Klioner}, {Lammers},
  {Lindegren}, {Luri}, {Mignard}, {Panem}, {Pourbaix}, {Randich}, {Sartoretti},
  {Soubiran}, {Walton}, {Arenou}, {Bailer-Jones}, {Bastian}, {Cropper},
  {Drimmel}, {Katz}, {Lattanzi}, {van Leeuwen}, {Bakker}, {Cacciari},
  {Casta{\~n}eda}, {De Angeli}, {Ducourant}, {Fabricius}, {Fouesneau},
  {Fr{\'e}mat}, {Guerra}, {Guerrier}, {Guiraud}, {Jean-Antoine Piccolo},
  {Masana}, {Messineo}, {Mowlavi}, {Nicolas}, {Nienartowicz}, {Pailler},
  {Panuzzo}, {Riclet}, {Roux}, {Seabroke}, {Sordo}, {Tanga}, {Th{\'e}venin},
  {Gracia-Abril}, {Portell}, {Teyssier}, {Altmann}, {Andrae}, {Bellas-Velidis},
  {Benson}, {Berthier}, {Blomme}, {Brugaletta}, {Burgess}, {Busso}, {Carry},
  {Cellino}, {Cheek}, {Clementini}, {Damerdji}, {Davidson}, {Delchambre},
  {Dell'Oro}, {Fern{\'a}ndez-Hern{\'a}ndez}, {Galluccio}, {Garc{\'\i}a-Lario},
  {Garcia-Reinaldos}, {Gonz{\'a}lez-N{\'u}{\~n}ez}, {Gosset}, {Haigron},
  {Halbwachs}, {Hambly}, {Harrison}, {Hatzidimitriou}, {Heiter},
  {Hern{\'a}ndez}, {Hestroffer}, {Hodgkin}, {Holl}, {Jan{\ss}en}, {Jevardat de
  Fombelle}, {Jordan}, {Krone-Martins}, {Lanzafame}, {L{\"o}ffler}, {Lorca},
  {Manteiga}, {Marchal}, {Marrese}, {Moitinho}, {Mora}, {Muinonen}, {Osborne},
  {Pancino}, {Pauwels}, {Petit}, {Recio-Blanco}, {Richards}, {Riello},
  {Rimoldini}, {Robin}, {Roegiers}, {Rybizki}, {Sarro}, {Siopis}, {Smith},
  {Sozzetti}, {Ulla}, {Utrilla}, {van Leeuwen}, {van Reeven}, {Abbas}, {Abreu
  Aramburu}, {Accart}, {Aerts}, {Aguado}, {Ajaj}, {Altavilla}, {{\'A}lvarez},
  {{\'A}lvarez Cid-Fuentes}, {Alves}, {Anderson}, {Anglada Varela}, {Antoja},
  {Audard}, {Baines}, {Baker}, {Balaguer-N{\'u}{\~n}ez}, {Balbinot}, {Balog},
  {Barache}, {Barbato}, {Barros}, {Barstow}, {Bartolom{\'e}}, {Bassilana},
  {Bauchet}, {Baudesson-Stella}, {Becciani}, {Bellazzini}, {Bernet}, {Bertone},
  {Bianchi}, {Blanco-Cuaresma}, {Boch}, {Bombrun}, {Bossini}, {Bouquillon},
  {Bragaglia}, {Bramante}, {Breedt}, {Bressan}, {Brouillet}, {Bucciarelli},
  {Burlacu}, {Busonero}, {Butkevich}, {Buzzi}, {Caffau}, {Cancelliere},
  {C{\'a}novas}, {Cantat-Gaudin}, {Carballo}, {Carlucci}, {Carnerero},
  {Carrasco}, {Casamiquela}, {Castellani}, {Castro-Ginard}, {Castro Sampol},
  {Chaoul}, {Charlot}, {Chemin}, {Chiavassa}, {Cioni}, {Comoretto}, {Cooper},
  {Cornez}, {Cowell}, {Crifo}, {Crosta}, {Crowley}, {Dafonte}, {Dapergolas},
  {David}, {David}, {de Laverny}, {De Luise}, {De March}, {De Ridder}, {de
  Souza}, {de Teodoro}, {de Torres}, {del Peloso}, {del Pozo}, {Delbo},
  {Delgado}, {Delgado}, {Delisle}, {Di Matteo}, {Diakite}, {Diener},
  {Distefano}, {Dolding}, {Eappachen}, {Edvardsson}, {Enke}, {Esquej}, {Fabre},
  {Fabrizio}, {Faigler}, {Fedorets}, {Fernique}, {Fienga}, {Figueras},
  {Fouron}, {Fragkoudi}, {Fraile}, {Franke}, {Gai}, {Garabato},
  {Garcia-Gutierrez}, {Garc{\'\i}a-Torres}, {Garofalo}, {Gavras}, {Gerlach},
  {Geyer}, {Giacobbe}, {Gilmore}, {Girona}, {Giuffrida}, {Gomel}, {Gomez},
  {Gonzalez-Santamaria}, {Gonz{\'a}lez-Vidal}, {Granvik},
  {Guti{\'e}rrez-S{\'a}nchez}, {Guy}, {Hauser}, {Haywood}, {Helmi}, {Hidalgo},
  {Hilger}, {H{\l}adczuk}, {Hobbs}, {Holland}, {Huckle}, {Jasniewicz},
  {Jonker}, {Juaristi Campillo}, {Julbe}, {Karbevska}, {Kervella}, {Khanna},
  {Kochoska}, {Kontizas}, {Kordopatis}, {Korn}, {Kostrzewa-Rutkowska},
  {Kruszy{\'n}ska}, {Lambert}, {Lanza}, {Lasne}, {Le Campion}, {Le Fustec},
  {Lebreton}, {Lebzelter}, {Leccia}, {Leclerc}, {Lecoeur-Taibi}, {Liao},
  {Licata}, {Lindstr{\o}m}, {Lister}, {Livanou}, {Lobel}, {Madrero Pardo},
  {Managau}, {Mann}, {Marchant}, {Marconi}, {Marcos Santos}, {Marinoni},
  {Marocco}, {Marshall}, {Martin Polo}, {Mart{\'\i}n-Fleitas}, {Masip},
  {Massari}, {Mastrobuono-Battisti}, {Mazeh}, {McMillan}, {Messina},
  {Michalik}, {Millar}, {Mints}, {Molina}, {Molinaro}, {Moln{\'a}r},
  {Montegriffo}, {Mor}, {Morbidelli}, {Morel}, {Morris}, {Mulone}, {Munoz},
  {Muraveva}, {Murphy}, {Musella}, {Noval}, {Ord{\'e}novic}, {Orr{\`u}},
  {Osinde}, {Pagani}, {Pagano}, {Palaversa}, {Palicio}, {Panahi}, {Pawlak},
  {Pe{\~n}alosa Esteller}, {Penttil{\"a}}, {Piersimoni}, {Pineau}, {Plachy},
  {Plum}, {Poggio}, {Poretti}, {Poujoulet}, {Pr{\v{s}}a}, {Pulone}, {Racero},
  {Ragaini}, {Rainer}, {Raiteri}, {Rambaux}, {Ramos}, {Ramos-Lerate}, {Re
  Fiorentin}, {Regibo}, {Reyl{\'e}}, {Ripepi}, {Riva}, {Rixon}, {Robichon},
  {Robin}, {Roelens}, {Rohrbasser}, {Romero-G{\'o}mez}, {Rowell}, {Royer},
  {Rybicki}, {Sadowski}, {Sagrist{\`a} Sell{\'e}s}, {Sahlmann}, {Salgado},
  {Salguero}, {Samaras}, {Sanchez Gimenez}, {Sanna}, {Santove{\~n}a},
  {Sarasso}, {Schultheis}, {Sciacca}, {Segol}, {Segovia}, {S{\'e}gransan},
  {Semeux}, {Shahaf}, {Siddiqui}, {Siebert}, {Siltala}, {Slezak}, {Smart},
  {Solano}, {Solitro}, {Souami}, {Souchay}, {Spagna}, {Spoto}, {Steele},
  {Steidelm{\"u}ller}, {Stephenson}, {S{\"u}veges}, {Szabados}, {Szegedi-Elek},
  {Taris}, {Tauran}, {Taylor}, {Teixeira}, {Thuillot}, {Tonello}, {Torra},
  {Torra}, {Turon}, {Unger}, {Vaillant}, {van Dillen}, {Vanel}, {Vecchiato},
  {Viala}, {Vicente}, {Voutsinas}, {Weiler}, {Wevers}, {Wyrzykowski}, {Yoldas},
  {Yvard}, {Zhao}, {Zorec}, {Zucker}, {Zurbach}, \&
  {Zwitter}}]{2021A&A...649A...1G}
{Gaia Collaboration}, {Brown}, A.~G.~A., {Vallenari}, A., {et~al.} 2021, \aap,
  649, A1

\bibitem[{{Gaia Collaboration} {et~al.}(2023){Gaia Collaboration}, {Creevey},
  {Sarro}, {Lobel}, {Pancino}, {Andrae}, {Smart}, {Clementini}, {Heiter},
  {Korn}, {Fouesneau}, {Fr{\'e}mat}, {De Angeli}, {Vallenari}, {Harrison},
  {Th{\'e}venin}, {Reyl{\'e}}, {Sordo}, {Garofalo}, {Brown}, {Eyer}, {Prusti},
  {de Bruijne}, {Arenou}, {Babusiaux}, {Biermann}, {Ducourant}, {Evans},
  {Guerra}, {Hutton}, {Jordi}, {Klioner}, {Lammers}, {Lindegren}, {Luri},
  {Mignard}, {Panem}, {Pourbaix}, {Randich}, {Sartoretti}, {Soubiran}, {Tanga},
  {Walton}, {Bailer-Jones}, {Bastian}, {Drimmel}, {Jansen}, {Katz}, {Lattanzi},
  {van Leeuwen}, {Bakker}, {Cacciari}, {Casta{\~n}eda}, {Fabricius},
  {Galluccio}, {Guerrier}, {Masana}, {Messineo}, {Mowlavi}, {Nicolas},
  {Nienartowicz}, {Pailler}, {Panuzzo}, {Riclet}, {Roux}, {Seabroke},
  {Gracia-Abril}, {Portell}, {Teyssier}, {Altmann}, {Audard}, {Bellas-Velidis},
  {Benson}, {Berthier}, {Blomme}, {Burgess}, {Busonero}, {Busso},
  {C{\'a}novas}, {Carry}, {Cellino}, {Cheek}, {Damerdji}, {Davidson}, {de
  Teodoro}, {Nu{\~n}ez Campos}, {Delchambre}, {Dell'Oro}, {Esquej},
  {Fern{\'a}ndez-Hern{\'a}ndez}, {Fraile}, {Garabato}, {Garc{\'\i}a-Lario},
  {Gosset}, {Haigron}, {Halbwachs}, {Hambly}, {Hern{\'a}ndez}, {Hestroffer},
  {Hodgkin}, {Holl}, {Jan{\ss}en}, {Jevardat de Fombelle}, {Jordan},
  {Krone-Martins}, {Lanzafame}, {L{\"o}ffler}, {Marchal}, {Marrese},
  {Moitinho}, {Muinonen}, {Osborne}, {Pauwels}, {Recio-Blanco}, {Riello},
  {Rimoldini}, {Roegiers}, {Rybizki}, {Siopis}, {Smith}, {Sozzetti}, {Utrilla},
  {van Leeuwen}, {Abbas}, {{\'A}brah{\'a}m}, {Abreu Aramburu}, {Aerts},
  {Aguado}, {Ajaj}, {Aldea-Montero}, {Altavilla}, {{\'A}lvarez}, {Alves},
  {Anders}, {Anderson}, {Anglada Varela}, {Antoja}, {Baines}, {Baker},
  {Balaguer-N{\'u}{\~n}ez}, {Balbinot}, {Balog}, {Barache}, {Barbato},
  {Barros}, {Barstow}, {Bartolom{\'e}}, {Bassilana}, {Bauchet}, {Becciani},
  {Bellazzini}, {Berihuete}, {Bernet}, {Bertone}, {Bianchi}, {Binnenfeld},
  {Blanco-Cuaresma}, {Boch}, {Bombrun}, {Bossini}, {Bouquillon}, {Bragaglia},
  {Bramante}, {Breedt}, {Bressan}, {Brouillet}, {Brugaletta}, {Bucciarelli},
  {Burlacu}, {Butkevich}, {Buzzi}, {Caffau}, {Cancelliere}, {Cantat-Gaudin},
  {Carballo}, {Carlucci}, {Carnerero}, {Carrasco}, {Casamiquela}, {Castellani},
  {Castro-Ginard}, {Chaoul}, {Charlot}, {Chemin}, {Chiaramida}, {Chiavassa},
  {Chornay}, {Comoretto}, {Contursi}, {Cooper}, {Cornez}, {Cowell}, {Crifo},
  {Cropper}, {Crosta}, {Crowley}, {Dafonte}, {Dapergolas}, {David}, {de
  Laverny}, {De Luise}, {De March}, {De Ridder}, {de Souza}, {de Torres}, {del
  Peloso}, {del Pozo}, {Delbo}, {Delgado}, {Delisle}, {Demouchy},
  {Dharmawardena}, {Di Matteo}, {Diakite}, {Diener}, {Distefano}, {Dolding},
  {Enke}, {Fabre}, {Fabrizio}, {Faigler}, {Fedorets}, {Fernique}, {Figueras},
  {Fournier}, {Fouron}, {Fragkoudi}, {Gai}, {Garcia-Gutierrez},
  {Garcia-Reinaldos}, {Garc{\'\i}a-Torres}, {Gavel}, {Gavras}, {Gerlach},
  {Geyer}, {Giacobbe}, {Gilmore}, {Girona}, {Giuffrida}, {Gomel}, {Gomez},
  {Gonz{\'a}lez-N{\'u}{\~n}ez}, {Gonz{\'a}lez-Santamar{\'\i}a},
  {Gonz{\'a}lez-Vidal}, {Granvik}, {Guillout}, {Guiraud},
  {Guti{\'e}rrez-S{\'a}nchez}, {Guy}, {Hatzidimitriou}, {Hauser}, {Haywood},
  {Helmer}, {Helmi}, {Hilger}, {Sarmiento}, {Hidalgo}, {H{\l}adczuk}, {Hobbs},
  {Holland}, {Huckle}, {Jardine}, {Jasniewicz}, {Jean-Antoine Piccolo},
  {Jim{\'e}nez-Arranz}, {Juaristi Campillo}, {Julbe}, {Karbevska}, {Kervella},
  {Khanna}, {Kordopatis}, {K{\'o}sp{\'a}l}, {Kostrzewa-Rutkowska},
  {Kruszy{\'n}ska}, {Kun}, {Laizeau}, {Lambert}, {Lanza}, {Lasne}, {Le
  Campion}, {Lebreton}, {Lebzelter}, {Leccia}, {Leclerc}, {Lecoeur-Taibi},
  {Liao}, {Licata}, {Lindstr{\o}m}, {Lister}, {Livanou}, {Lorca}, {Loup},
  {Madrero Pardo}, {Magdaleno Romeo}, {Managau}, {Mann}, {Manteiga},
  {Marchant}, {Marconi}, {Marcos}, {Marcos Santos}, {Mar{\'\i}n Pina},
  {Marinoni}, {Marocco}, {Marshall}, {Martin Polo}, {Mart{\'\i}n-Fleitas},
  {Marton}, {Mary}, {Masip}, {Massari}, {Mastrobuono-Battisti}, {Mazeh},
  {McMillan}, {Messina}, {Michalik}, {Millar}, {Mints}, {Molina}, {Molinaro},
  {Moln{\'a}r}, {Monari}, {Mongui{\'o}}, {Montegriffo}, {Montero}, {Mor},
  {Mora}, {Morbidelli}, {Morel}, {Morris}, {Muraveva}, {Murphy}, {Musella},
  {Nagy}, {Noval}, {Oca{\~n}a}, {Ogden}, {Ordenovic}, {Osinde}, {Pagani},
  {Pagano}, {Palaversa}, {Palicio}, {Pallas-Quintela}, {Panahi},
  {Payne-Wardenaar}, {Pe{\~n}alosa Esteller}, {Penttil{\"a}}, {Pichon},
  {Piersimoni}, {Pineau}, {Plachy}, {Plum}, {Poggio}, {Pr{\v{s}}a}, {Pulone},
  {Racero}, {Ragaini}, {Rainer}, {Raiteri}, {Ramos}, {Ramos-Lerate}, {Re
  Fiorentin}, {Regibo}, {Richards}, {Rios Diaz}, {Ripepi}, {Riva}, {Rix},
  {Rixon}, {Robichon}, {Robin}, {Robin}, {Roelens}, {Rogues}, {Rohrbasser},
  {Romero-G{\'o}mez}, {Rowell}, {Royer}, {Ruz Mieres}, {Rybicki}, {Sadowski},
  {S{\'a}ez N{\'u}{\~n}ez}, {Sagrist{\`a} Sell{\'e}s}, {Sahlmann}, {Salguero},
  {Samaras}, {Sanchez Gimenez}, {Sanna}, {Santove{\~n}a}, {Sarasso},
  {Schultheis}, {Sciacca}, {Segol}, {Segovia}, {S{\'e}gransan}, {Semeux},
  {Shahaf}, {Siddiqui}, {Siebert}, {Siltala}, {Silvelo}, {Slezak}, {Slezak},
  {Snaith}, {Solano}, {Solitro}, {Souami}, {Souchay}, {Spagna}, {Spina},
  {Spoto}, {Steele}, {Steidelm{\"u}ller}, {Stephenson}, {S{\"u}veges},
  {Surdej}, {Szabados}, {Szegedi-Elek}, {Taris}, {Taylor}, {Teixeira},
  {Tolomei}, {Tonello}, {Torra}, {Torra}, {Torralba Elipe}, {Trabucchi},
  {Tsounis}, {Turon}, {Ulla}, {Unger}, {Vaillant}, {van Dillen}, {van Reeven},
  {Vanel}, {Vecchiato}, {Viala}, {Vicente}, {Voutsinas}, {Weiler}, {Wevers},
  {Wyrzykowski}, {Yoldas}, {Yvard}, {Zhao}, {Zorec}, {Zucker}, \&
  {Zwitter}}]{2023A&A...674A..39G}
{Gaia Collaboration}, {Creevey}, O.~L., {Sarro}, L.~M., {et~al.} 2023, \aap,
  674, A39

\bibitem[{{Ivezi{\'c}} {et~al.}(2012){Ivezi{\'c}}, {Beers}, \&
  {Juri{\'c}}}]{2012ARA&A..50..251I}
{Ivezi{\'c}}, {\v{Z}}., {Beers}, T.~C., \& {Juri{\'c}}, M. 2012, \araa, 50, 251

\bibitem[{{Ivezi{\'c}} {et~al.}(2019){Ivezi{\'c}}, {Kahn}, {Tyson}, {Abel},
  {Acosta}, {Allsman}, {Alonso}, {AlSayyad}, {Anderson}, {Andrew}, {Angel},
  {Angeli}, {Ansari}, {Antilogus}, {Araujo}, {Armstrong}, {Arndt}, {Astier},
  {Aubourg}, {Auza}, {Axelrod}, {Bard}, {Barr}, {Barrau}, {Bartlett}, {Bauer},
  {Bauman}, {Baumont}, {Bechtol}, {Bechtol}, {Becker}, {Becla}, {Beldica},
  {Bellavia}, {Bianco}, {Biswas}, {Blanc}, {Blazek}, {Blandford}, {Bloom},
  {Bogart}, {Bond}, {Booth}, {Borgland}, {Borne}, {Bosch}, {Boutigny},
  {Brackett}, {Bradshaw}, {Brandt}, {Brown}, {Bullock}, {Burchat}, {Burke},
  {Cagnoli}, {Calabrese}, {Callahan}, {Callen}, {Carlin}, {Carlson},
  {Chandrasekharan}, {Charles-Emerson}, {Chesley}, {Cheu}, {Chiang}, {Chiang},
  {Chirino}, {Chow}, {Ciardi}, {Claver}, {Cohen-Tanugi}, {Cockrum}, {Coles},
  {Connolly}, {Cook}, {Cooray}, {Covey}, {Cribbs}, {Cui}, {Cutri}, {Daly},
  {Daniel}, {Daruich}, {Daubard}, {Daues}, {Dawson}, {Delgado}, {Dellapenna},
  {de Peyster}, {de Val-Borro}, {Digel}, {Doherty}, {Dubois},
  {Dubois-Felsmann}, {Durech}, {Economou}, {Eifler}, {Eracleous}, {Emmons},
  {Fausti Neto}, {Ferguson}, {Figueroa}, {Fisher-Levine}, {Focke}, {Foss},
  {Frank}, {Freemon}, {Gangler}, {Gawiser}, {Geary}, {Gee}, {Geha}, {Gessner},
  {Gibson}, {Gilmore}, {Glanzman}, {Glick}, {Goldina}, {Goldstein}, {Goodenow},
  {Graham}, {Gressler}, {Gris}, {Guy}, {Guyonnet}, {Haller}, {Harris},
  {Hascall}, {Haupt}, {Hernandez}, {Herrmann}, {Hileman}, {Hoblitt}, {Hodgson},
  {Hogan}, {Howard}, {Huang}, {Huffer}, {Ingraham}, {Innes}, {Jacoby}, {Jain},
  {Jammes}, {Jee}, {Jenness}, {Jernigan}, {Jevremovi{\'c}}, {Johns}, {Johnson},
  {Johnson}, {Jones}, {Juramy-Gilles}, {Juri{\'c}}, {Kalirai}, {Kallivayalil},
  {Kalmbach}, {Kantor}, {Karst}, {Kasliwal}, {Kelly}, {Kessler}, {Kinnison},
  {Kirkby}, {Knox}, {Kotov}, {Krabbendam}, {Krughoff}, {Kub{\'a}nek},
  {Kuczewski}, {Kulkarni}, {Ku}, {Kurita}, {Lage}, {Lambert}, {Lange},
  {Langton}, {Le Guillou}, {Levine}, {Liang}, {Lim}, {Lintott}, {Long},
  {Lopez}, {Lotz}, {Lupton}, {Lust}, {MacArthur}, {Mahabal}, {Mandelbaum},
  {Markiewicz}, {Marsh}, {Marshall}, {Marshall}, {May}, {McKercher}, {McQueen},
  {Meyers}, {Migliore}, {Miller}, {Mills}, {Miraval}, {Moeyens}, {Moolekamp},
  {Monet}, {Moniez}, {Monkewitz}, {Montgomery}, {Morrison}, {Mueller},
  {Muller}, {Mu{\~n}oz Arancibia}, {Neill}, {Newbry}, {Nief}, {Nomerotski},
  {Nordby}, {O'Connor}, {Oliver}, {Olivier}, {Olsen}, {O'Mullane}, {Ortiz},
  {Osier}, {Owen}, {Pain}, {Palecek}, {Parejko}, {Parsons}, {Pease},
  {Peterson}, {Peterson}, {Petravick}, {Libby Petrick}, {Petry},
  {Pierfederici}, {Pietrowicz}, {Pike}, {Pinto}, {Plante}, {Plate}, {Plutchak},
  {Price}, {Prouza}, {Radeka}, {Rajagopal}, {Rasmussen}, {Regnault}, {Reil},
  {Reiss}, {Reuter}, {Ridgway}, {Riot}, {Ritz}, {Robinson}, {Roby}, {Roodman},
  {Rosing}, {Roucelle}, {Rumore}, {Russo}, {Saha}, {Sassolas}, {Schalk},
  {Schellart}, {Schindler}, {Schmidt}, {Schneider}, {Schneider}, {Schoening},
  {Schumacher}, {Schwamb}, {Sebag}, {Selvy}, {Sembroski}, {Seppala}, {Serio},
  {Serrano}, {Shaw}, {Shipsey}, {Sick}, {Silvestri}, {Slater}, {Smith},
  {Smith}, {Sobhani}, {Soldahl}, {Storrie-Lombardi}, {Stover}, {Strauss},
  {Street}, {Stubbs}, {Sullivan}, {Sweeney}, {Swinbank}, {Szalay}, {Takacs},
  {Tether}, {Thaler}, {Thayer}, {Thomas}, {Thornton}, {Thukral}, {Tice},
  {Trilling}, {Turri}, {Van Berg}, {Vanden Berk}, {Vetter}, {Virieux},
  {Vucina}, {Wahl}, {Walkowicz}, {Walsh}, {Walter}, {Wang}, {Wang}, {Warner},
  {Wiecha}, {Willman}, {Winters}, {Wittman}, {Wolff}, {Wood-Vasey}, {Wu},
  {Xin}, {Yoachim}, \& {Zhan}}]{2019ApJ...873..111I}
{Ivezi{\'c}}, {\v{Z}}., {Kahn}, S.~M., {Tyson}, J.~A., {et~al.} 2019, \apj,
  873, 111

\bibitem[{{Ivezi{\'c}} {et~al.}(2008){Ivezi{\'c}}, {Sesar}, {Juri{\'c}},
  {Bond}, {Dalcanton}, {Rockosi}, {Yanny}, {Newberg}, {Beers}, {Allende
  Prieto}, {Wilhelm}, {Lee}, {Sivarani}, {Norris}, {Bailer-Jones}, {Re
  Fiorentin}, {Schlegel}, {Uomoto}, {Lupton}, {Knapp}, {Gunn}, {Covey}, {Allyn
  Smith}, {Miknaitis}, {Doi}, {Tanaka}, {Fukugita}, {Kent}, {Finkbeiner},
  {Munn}, {Pier}, {Quinn}, {Hawley}, {Anderson}, {Kiuchi}, {Chen}, {Bushong},
  {Sohi}, {Haggard}, {Kimball}, {Barentine}, {Brewington}, {Harvanek},
  {Kleinman}, {Krzesinski}, {Long}, {Nitta}, {Snedden}, {Lee}, {Harris},
  {Brinkmann}, {Schneider}, \& {York}}]{2008ApJ...684..287I}
{Ivezi{\'c}}, {\v{Z}}., {Sesar}, B., {Juri{\'c}}, M., {et~al.} 2008, \apj, 684,
  287

\bibitem[{{Juri{\'c}} {et~al.}(2008){Juri{\'c}}, {Ivezi{\'c}}, {Brooks},
  {Lupton}, {Schlegel}, {Finkbeiner}, {Padmanabhan}, {Bond}, {Sesar},
  {Rockosi}, {Knapp}, {Gunn}, {Sumi}, {Schneider}, {Barentine}, {Brewington},
  {Brinkmann}, {Fukugita}, {Harvanek}, {Kleinman}, {Krzesinski}, {Long},
  {Neilsen}, {Nitta}, {Snedden}, \& {York}}]{2008ApJ...673..864J}
{Juri{\'c}}, M., {Ivezi{\'c}}, {\v{Z}}., {Brooks}, A., {et~al.} 2008, \apj,
  673, 864

\bibitem[{{Li} \& {Widrow}(2023)}]{2023MNRAS.520.3329L}
{Li}, H. \& {Widrow}, L.~M. 2023, \mnras, 520, 3329

\bibitem[{{Loebman} {et~al.}(2014){Loebman}, {Ivezi{\'c}}, {Quinn}, {Bovy},
  {Christensen}, {Juri{\'c}}, {Ro{\v{s}}kar}, {Brooks}, \&
  {Governato}}]{2014ApJ...794..151L}
{Loebman}, S.~R., {Ivezi{\'c}}, {\v{Z}}., {Quinn}, T.~R., {et~al.} 2014, \apj,
  794, 151

\bibitem[{{Loebman} {et~al.}(2012){Loebman}, {Ivezi{\'c}}, {Quinn},
  {Governato}, {Brooks}, {Christensen}, \& {Juri{\'c}}}]{2012ApJ...758L..23L}
{Loebman}, S.~R., {Ivezi{\'c}}, {\v{Z}}., {Quinn}, T.~R., {et~al.} 2012, \apjl,
  758, L23

\bibitem[{{Loebman} {et~al.}(2011){Loebman}, {Ro{\v{s}}kar}, {Debattista},
  {Ivezi{\'c}}, {Quinn}, \& {Wadsley}}]{2011ApJ...737....8L}
{Loebman}, S.~R., {Ro{\v{s}}kar}, R., {Debattista}, V.~P., {et~al.} 2011, \apj,
  737, 8

\bibitem[{{Salomon} {et~al.}(2020){Salomon}, {Bienaym{\'e}}, {Reyl{\'e}},
  {Robin}, \& {Famaey}}]{2020A&A...643A..75S}
{Salomon}, J.-B., {Bienaym{\'e}}, O., {Reyl{\'e}}, C., {Robin}, A.~C., \&
  {Famaey}, B. 2020, \aap, 643, A75

\end{thebibliography}

\begin{appendix}
\section{Validation of proper motion systematic and random uncertainties using quasars}
We tested \textit{Gaia's} proper motions and their uncertainties using
spectroscopically confirmed quasars from SDSS Data Release 7. There
are $\sim$367,000 SDSS quasars with \textit{Gaia's} non-negative proper motion errors.
% (note that columns GAIA\_PM\_RA\_ERR and GAIA\_PM\_DEC\_ERR actually list inverse variance).
Their median  proper motion per
coordinate is about 0.01 mas/yr (indicating no substantial systematic
measurement errors) and the median proper motion magnitude
is about 1.1 mas/yr (indicating typical measurement uncertainty; the
median magnitude of this sample is $G\sim$20; for the FGKM sample
analyzed here, with $G<18$, the proper motion uncertainties are
$<0.15$ mas/yr). \\

We have verified that the width of  proper motion per
coordinate normalized by reported uncertainties (\ie, the width of corresponding
$\chi$ distributions) is 1.07 and 1.09, demonstrating \textit{Gaia's} reliable
estimates of measurement uncertainties. We did not find any significant variation in the median quasar proper motion per
coordinate with position on the sky. The only ``interesting feature''
in the data is an increased scatter of proper motion per
coordinate measurements in the so-called SDSS Stripe 82 region by
about 50\% compared to the rest of the SDSS sky. This effect is easily understood as being due to deeper quasar
sample in that region (due to details in the SDSS spectroscopic target
selection) and the increase of \textit{Gaia's} measurement uncertainties with
magnitude (and verified through no substantial increase in the
corresponding $\chi$ distributions). 

\end{appendix}

\end{document}